%% file: ms.tex
\documentclass[twocolumn]{aastex631}
\usepackage{longtable}
\usepackage{soul}
\usepackage{amsmath}
\usepackage{hyperref}
\usepackage{cleveref}
\usepackage{graphicx}
\usepackage{tabularx}
\usepackage{threeparttable}

\newcommand{\beq}{\begin{equation}}
\newcommand{\eeq}{\end{equation}}
\providecommand{\sw}[1]{\texttt{#1}}

\def\sn{SN~2025qe}
\newcommand{\Msun}{\ensuremath{M_{\odot}}}

\shorttitle{SN~2025qe: a low luminosity thermonuclear explosion}
\shortauthors{Das et al.}

\begin{document}
\title{SN 2025qe: A case study of diversity in low-luminosity thermonuclear explosions}

\correspondingauthor{Hrishav Das}
\email{hrishav.das@iiap.res.in, hrishav.hd@gmail.com}

\author[0009-0007-9727-7792]{Hrishav Das}
\affiliation{Indian Institute of Astrophysics, Koramangala 2nd Block, Bangalore 560034, India}
\affiliation{Pondicherry University, R.V. Nagar, Kalapet, 605014, Puducherry, India}

\author[0000-0002-7708-3831]{Anirban Dutta}
\affiliation{Graduate Institute of Astronomy, National Central University, 300 Zhongda Road, 32001 Zhongli, Taiwan}

\author[0000-0002-6688-0800]{Devendra K. Sahu}
\affiliation{Indian Institute of Astrophysics, Koramangala 2nd Block, Bangalore 560034, India}

\author[0000-0003-3533-7183]{G.C. Anupama}
\affiliation{Indian Institute of Astrophysics, Koramangala 2nd Block, Bangalore 560034, India}

\author[0009-0001-7833-5703]{Prathamesh S. Kadam}
\affiliation{Department of Physics, Sardar Vallabhbhai National Institute of Technology, Surat, 395007, Gujarat, India}

\author[0000-0002-7942-8477]{Vishwajeet Swain}
\affiliation{Department of Physics, Indian Institute of Technology Bombay, Powai, 400076, India}

\begin{abstract}
We present optical photometric and spectroscopic observations of SN~2025qe, a Type Iax supernova in IC~0529, monitored from $-10$ to $+150$ rest-frame days relative to $g'$-band maximum using the GROWTH-India and the Himalayan Chandra Telescopes. SN~2025qe reaches a peak $g'$-band absolute magnitude of $M_{g'} \approx -16.16 \pm 0.15$~mag, placing it in the sparsely sampled luminosity gap between the brightest and faintest members of the SNe~Iax class. The multiband light curves show pronounced wavelength-dependent broadening at late times, with the redder bands remaining brighter than in other SNe~Iax. Semi-analytical (Arnett) modeling of the pseudobolometric light curve yields a synthesized $^{56}$Ni mass of $0.02\,M_\odot$, an ejecta mass of $0.45\,M_\odot$, and a kinetic energy of
$7.0\times 10^{49}$~erg. Three-dimensional pure-deflagration models of Chandrasekhar-mass carbon-oxygen white dwarfs reproduce the peak luminosity and $^{56}$Ni yield but underpredict the slow, red post-maximum decline; models including energy input from a bound remnant better match the late-time evolution, especially in the red bands. We model early-time spectra using \textsc{tardis} with a modified deflagration model and identify late-time spectral features using \texttt{syn++}. We qualitatively try to explain the persistent red-band flux with a two-component ejecta structure comprising a dense inner region, possibly including a bound remnant, and a density-enhanced outer shell that sustains redder emission. Together, the photometric and spectroscopic properties of SN~2025qe highlight the diversity of ejecta structure and evolution among SNe~Iax with similar peak luminosities, emphasizing the need for detailed observations and modeling of their ejecta.
\end{abstract}

\keywords{Supernovae (1668) --- Type Ia supernovae (1728) --- White dwarf stars (1799)}

\section{Introduction} \label{sec:intro}

Type Ia supernovae (SNe Ia) are widely understood to result from the thermonuclear disruption of a carbon-oxygen (CO) white dwarf (WD) in a binary system once the WD approaches the Chandrasekhar mass and undergoes runaway nuclear burning \citep{1960ApJ...132..565H}. Most normal SNe Ia form a homogeneous class, often termed `Branch-normal' SNe \citep{2006PASP..118..560B}. Their peak luminosity strongly correlate with the shape of their light curve \citep{1993ApJ...413L.105P,1999AJ....118.1766P}, a relationship governed by the amount of radioactive $^{56}$Ni synthesized during the explosion \citep{1982ApJ...253..785A}. This makes them vital cosmological tools, as they can be used as standardizable candles to measure cosmic distances. However, a growing number of peculiar thermonuclear explosions, though still assumed to originate from CO WDs, deviate significantly from this standard correlation and exhibit unusual observables compared to normal SNe Ia.

One significant subclass of these peculiar explosions is the Type Iax Supernovae (SNe Iax), also known as `2002cx-like' SNe \citep{2003PASP..115..453L}. SNe Iax are considered the peculiar, fainter cousins of SNe Ia, characterized by comparatively lower explosion energies, luminosities, and ejecta velocities \citep{2013ApJ...767...57F,2017hsn..book..375J}.
Although the exact rate of SNe Iax is uncertain, estimates suggest they are frequent, potentially occurring at 15–30\% of the normal SN Ia rate \citep{2013ApJ...767...57F,2022MNRAS.511.2708S}, making the group perhaps the most numerous Ia subclass.

These SNe are frequently associated with late-type, star-forming host galaxies \citep{2013ApJ...767...57F,2013MNRAS.434..527L,2024MNRAS.533.3517Q}, although rare exceptions are known. Environmental studies using photometric and integral-field data show that these explosions tend to occur in regions dominated by young stellar populations, implying comparatively short delay times for at least a substantial fraction of the population. These trends support single-degenerate progenitor channels in which a CO WD accretes from a helium-rich donor \citep{2015A&A...574A..12L,2018MNRAS.473.1359L}. Recent analyses of explosion-site stellar populations further strengthen this picture. Studies using resolved \textit{HST} photometry find that SNe Iax typically arise from environments with sub-solar metallicities and stellar ages of only $\sim$40–100 Myr, consistent with young, helium-star donor systems rather than double-degenerate or core-collapse channels \citep{2020MNRAS.493..986T}. Recent theoretical work further reinforces this interpretation by showing that single-degenerate explosions can naturally reproduce key observed features of SNe Iax, including low kinetic energies, incomplete burning, and the survival of a bound remnant \citep{2025arXiv250716907M}. Direct evidence for progenitor systems has emerged in a handful of cases. Pre-explosion imaging of SN 2012Z revealed a luminous blue source consistent with a helium-star companion \citep{2014Natur.512...54M}, and later observations confirmed that both the companion and a bound remnant likely survived the explosion \citep{McCully2022,schwab2025remarkablelatetimefluxexcess}. Similar signatures have been identified for several other SNe Iax, including SN 2008ha and SN 2014dt \citep{2014ApJ...792...29F,2015ApJ...798L..37F,2023MNRAS.525.1210M}. 

SNe Iax show remarkable diversity in their observational properties. Their peak absolute magnitudes span a broad range, from extremely faint events such as SN 2021fcg ($M_{g} = -11.7$ mag; \citealt{Karambelkar_2021}) to nearly as bright as normal SNe Ia such as SN 2012Z ($M_{g} = -18.4$ mag; \citealt{2015A&A...573A...2S}). They exhibit comparatively modest kinetic energies, reflected in photospheric velocities of only a few thousand kilometres per second \citep{2017hsn..book..375J}, substantially lower than those measured for typical SNe Ia, which show expansion velocities well above 10,000 km s$^{-1}$ \citep{2009ApJ...699L.139W}. The rise times of SNe Iax are often shorter than those of SNe Ia when pre-maximum data are available \citep{2016A&A...589A..89M,2017A&A...601A..62M}. Several studies have examined whether the photometric diversity of SNe Iax follows underlying trends. Although some works have suggested a broad tendency for brighter events to exhibit higher expansion velocities \citep{2010ApJ...720..704M}, there remains uncertainty about whether this reflects an actual physical correlation, as some objects deviate significantly from the proposed relation, such as SN 2014ck and SN 2009ku \citep{2016MNRAS.459.1018T}. Similar attempts to link peak luminosity with rise time or with post-maximum decline rate have suggested possible trends. However, the data show considerable scatter, and no reliable, universal relation has yet been established for this class of objects \citep{2016A&A...589A..89M}.

Spectroscopically, SNe Iax exhibit distinctive characteristics from the earliest epochs. Pre-maximum spectra are dominated by strong Fe\,{\sc iii} and Fe\,{\sc ii} features, with unusually weak Si\,{\sc ii} absorption, similar to what is observed in 91T-like SNe Ia \citep{2007PASP..119..360P,2013ApJ...767...57F}. Despite this similarity, SNe Iax possess significantly lower expansion velocities than 91T-like SNe Ia \citep{2005ApJ...623.1011B}. Carbon signatures also vary across the class: high-luminosity Iax events typically lack strong C\,{\sc ii} absorption, whereas several faint and intermediate-luminosity SNe Iax display clear detections of unburned carbon \citep{2009AJ....138..376F,2014A&A...561A.146S,2022MNRAS.511.2708S,2023ApJ...953...93S}. At late epochs, SNe Iax do not transition cleanly into the nebular phase; instead, their spectra retain a mixture of permitted and forbidden lines for hundreds of days, a behavior attributed to the presence of a long-lived, optically thick wind powered by a surviving bound remnant \citep{2016MNRAS.461..433F,2022ApJ...941...15M,2023ApJ...951...67C}. In addition to these spectroscopic signatures, several studies have proposed possible direct detections of surviving remnants. A notable example is the point source discovered at the position of SN 2008ha in late-time \textit{HST} imaging, interpreted as either a companion star or a partially burnt remnant \citep{2014ApJ...792...29F}. Further support comes from the identification of high-proper-motion, low-mass Galactic WDs \citep{2017Sci...357..680V,2019MNRAS.489.1489R}, which have been proposed as bound remnants ejected from their progenitor systems following weak deflagration explosions.

The pronounced diversity in luminosity and other observables among SNe Iax raises the possibility that more than one explosion pathway may be required to explain the 2002cx-like SNe population. Although multiple explosion scenarios have been proposed to explain the heterogeneous properties of SNe Iax, weak deflagrations of Chandrasekhar-mass CO WDs, which fail to completely unbind the progenitor and leave behind a bound remnant \citep{2013MNRAS.429.2287K,2014MNRAS.438.1762F}, remain the most successful in reproducing many of the observed features of the brighter members of the class (M$_{r}$$\leq$ -17.1 mag, \citealt{2023ApJ...953...93S}). Recent hydrodynamic simulations have explored a wider range of initial conditions, including the ignition location, central density, metallicity, and progenitor composition. They show that pure deflagrations in Chandrasekhar-mass carbon–oxygen WDs can, in principle, produce a broad range of luminosities and kinetic energies similar to those observed in bright and intermediate-luminosity SNe \citep[][see their parameter study]{2022A&A...658A.179L}. For fainter SNe Iax (M$_{r}$$\geq$ -14.64 mag, \citealt{2023ApJ...953...93S}), weak deflagrations of hybrid cabon-oxygen-neon (CONe) WDs have been proposed as an alternative \citep{2014ApJ...789L..45M,2015MNRAS.450.3045K}. However, challenges remain for pure deflagration models, as synthetic light curves sometimes decline too rapidly after maximum, potentially indicating insufficient ejecta masses \citep{2026ApJ...999...10D}. Other possible pathways include CO+ONe WD mergers \citep{2018ApJ...869..140K} and WD-compact object interactions \citep{2022MNRAS.510.3758B}. To better understand the diverse observables in SNe Iax in terms of the explosion physics, more realistic hydrodynamic models and detailed radiative-transfer calculations are needed.

A significant obstacle to confirming a unified origin for the SNe Iax class is the lack of detailed observations for moderately luminous objects that fall between the high-luminosity events (like SN 2005hk, \citealt{2008ApJ...680..580S}) and the extremely faint events (like SN 2008ha, \citealt{2009AJ....138..376F}). Investigating SNe Iax that populate this luminosity gap is important for understanding whether their physical and chemical properties vary continuously with peak luminosity, which would strongly support the hypothesis that they share a similar origin, such as the pure deflagration of a WD. Recent studies have focused on SNe that begin to bridge this gap, such as SN 2019muj and SN 2024pxl. SN 2019muj was found to be a moderate luminosity object ($M_g \approx -16.48$ mag, \citealt{2021MNRAS.501.1078B}), providing a unique opportunity to link the extremely low-luminosity SNe Iax to the brighter, well-studied members. Similarly, the nearby SN 2024pxl, with a peak absolute magnitude of $M_V \approx -16.60$ mag \citep{2025arXiv250502944K}, also exhibited photometric and spectroscopic properties intermediate between the high-luminosity and low-luminosity groups, further establishing a continuous link across the class.

\begin{figure}
	\begin{center}
		\resizebox{\hsize}{!}{\includegraphics{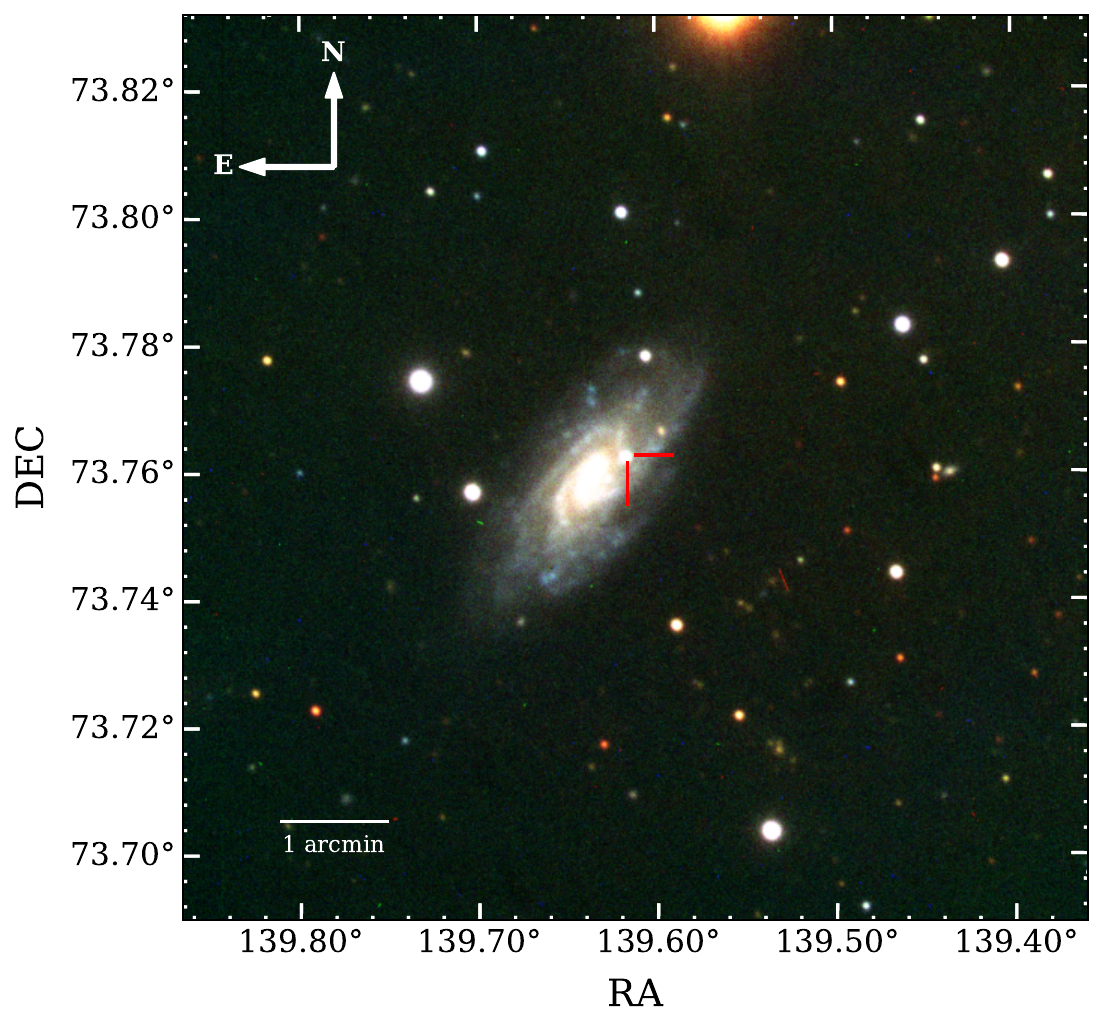}}
	\end{center}
	\caption{$u'g'r'i'z'$ composite image of SN 2025qe obtained with the 2-m Himalayan Chandra Telescope (HCT), showing a field of view of approximately $8.5' \times 8.5'$. The position of SN 2025qe is marked by red crosshairs, the orientation is indicated by the north and east arrows, and the scale bar corresponds to 1 arcmin.}
	\label{fig:SN_field}
\end{figure}

In this paper, we present extensive photometric and spectroscopic observations of \sn, which was first detected by the Zwicky Transient Facility (ZTF) on 2025 January 18 at 06:17:58 UT (JD = 2460693.76), with a $g_{\rm ZTF}$-band magnitude of 18.67 mag, in a nearby SA(s)c galaxy \citep{1991rc3..book.....D},  IC 0529 \citep{2025TNSTR.242....1S}, at a redshift of $z = 0.0075$. Its location within its host galaxy is shown in Figure \ref{fig:SN_field}. The transient, designated ZTF25aacerkv, was subsequently classified as a Type Iax supernova by the GOTO collaboration using a spectrum obtained on 2025 January 20 \citep{2025TNSCR.278....1P}. 
By analyzing the photometric and spectroscopic evolution of \sn, we aim to constrain its physical parameters, test existing hydrodynamic models, and assess whether the observed diversity in SNe Iax can be explained with a single underlying explosion mechanism.

The observations and data reduction procedures are presented in Section \ref{observation and data reduction}. Section \ref{distance_extinction_explosion_epoch} discusses the adopted distance, extinction, and explosion epoch. Section \ref{light_curve} examines the photometric evolution and compares synthetic light curves from pure deflagration models with the data. The spectroscopic analysis, including radiative-transfer modeling with TARDIS, is described in Section \ref{spectral analysis}. The principal results and conclusions are discussed in Section \ref{discussion}.

\begin{figure*}[hbt!]
	\begin{center}
		
        \includegraphics[width=\textwidth]{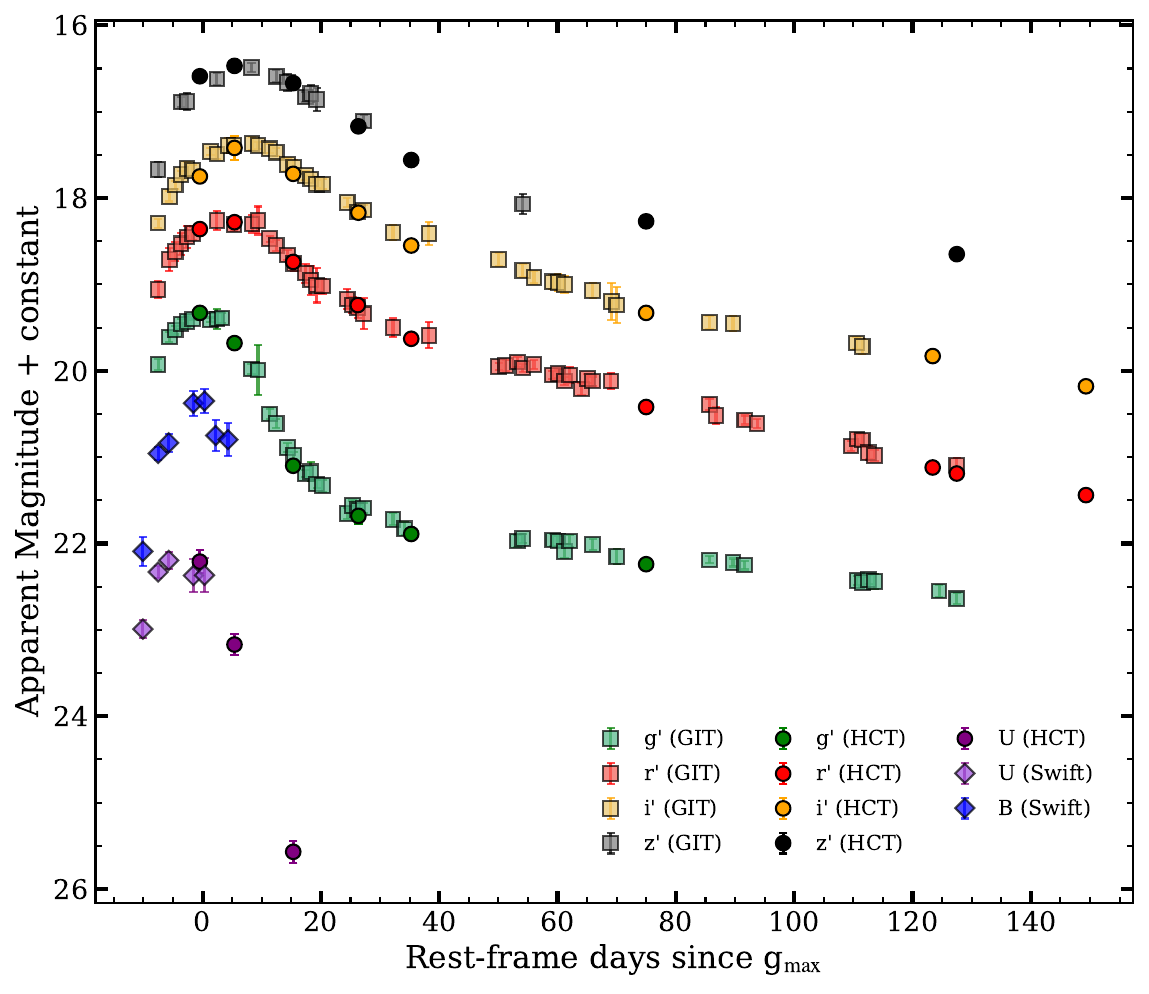}
	\end{center}
	\caption{Light curve evolution of SN~2025qe in $UBg'r'i'z'$ bands. The y-axis shows the apparent magnitudes in the respective bands, shifted by a constant for a clear representation. The light curve in each band is plotted with its corresponding error bars. The magnitudes in $U$ and $B$ bands are in the \textit{Vega} system and the magnitudes in $g'r'i'z'$ bands are plotted in the AB system. The data used to generate this figure are provided as data behind the figure.}
	\label{fig:SN_2025qe_light_curve}
\end{figure*}

\section{Observations and Data Reduction}
\label{observation and data reduction}

\subsection{Photometry}

\subsubsection{GIT} \label{GIT}
The imaging observations of SN 2025qe, from 2025-01-21 (JD 2460697.2) to 2025-06-06 (JD 2460833.2), were carried out using GROWTH-India Telescope (GIT) \citep{2022AJ....164...90K}, located at the Indian Astronomical Observatory (IAO) in Hanle, Ladakh. GIT is a 0.7-meter wide-field, fully robotic telescope designed specifically for studying astrophysical transients. The observations were taken with the STC-428 CMOS Imaging Camera (SBIG) in SDSS $g'$, $r'$, $ i'$, and $z'$ filters.

Data were downloaded and processed in real time using the GIT data reduction pipeline. This included basic reduction, astrometry, and point-spread function (PSF) photometry, following the procedures outlined in \citet{2022AJ....164...90K}. We used the \sw{ZOGY} algorithm in a \texttt{Python} pipeline to perform image subtraction on all images with the PS1 templates. The pipeline then applied PSF fit photometry to the subtracted images to obtain magnitudes, which were calibrated by cross-matching them against the PS1 DR1 catalog \citep{2016arXiv161205560C} through Vizier.

\subsubsection{HCT} \label{HCT}
SN 2025qe was also observed in the SDSS $u'$, $g'$, $r'$, $i'$, and $z'$ filters using the 2-meter Himalayan Chandra Telescope (HCT) \citep{2010ASInC...1..193P}, at IAO,  from 2025-01-28 (JD 2460704.3) to 2025-06-28 (JD 2460855.2). 
The Himalayan Faint Object Spectrograph Camera (HFOSC) was used for  optical imaging and low-to-medium resolution spectroscopy.  

For the $g'$, $r'$, $i'$, and $z'$ filters, the data were processed following the method described in Section \S\ref{GIT}. For the $u'$-band data, since no suitable SDSS template was available, a template image was obtained on 2025-11-21, after the supernova had faded below the limiting magnitude of the telescope. Image subtraction was then performed using this template. On the same night, standard fields were also observed in the $u'$-band and calibrated by us. These calibrated standard fields were then used to determine calibrated magnitudes for a set of stars in the supernova field, which served as secondary standards for the photometric calibration of the science images.

\subsubsection{Swift/UVOT} \label{Swift/UVOT}

SN 2025qe was observed with the UVOT instrument of {\it Swift} Observatory \citep{2005roming} aboard the Neil Gehrels Swift Observatory \citep{2004gehrels}. The processed images were retrieved from the Swift archive and were available in the $V$, $B$, $U$, $uvw1$ (2600 \AA), $uvm2$ (2246 \AA), and $uvw2$ (1928 \AA) filters. Photometry was performed on the processed images using the {\it uvotsource} task \citep{poole2008MNRAS.383..627P}.

In the $V$, $uvw1$, $uvm2$, and $uvw2$ bands, the source counts in all epochs were comparable to the background level, and therefore no reliable photometric measurements could be obtained. In the $U$ and $B$ bands, magnitudes were measured for those epochs where the source counts were significantly above the background.  In these bands, the underlying host-galaxy counts in the vicinity of the supernova were much lower than the source counts, and since no suitable UVOT template images were available, the background was estimated using multiple source-free regions distributed around the transient location to account for the local sky contribution. The area-weighted mean of these regions was then subtracted from source counts to get the corresponding magnitudes.

\subsection{Spectroscopy}
We obtained the spectra of \sn~ over a period of 119 days, between 2025-01-26 and 2025-06-28, to monitor its spectroscopic evolution. The data were acquired using HFOSC instrument mounted on HCT. The instrument is equipped with grisms Gr7 and Gr8, which together provide wavelength coverage over the optical range of 3500–9100 \AA.

The observed two-dimensional spectra were reduced following standard procedures in \texttt{IRAF}, which included bias subtraction, flat-field correction, wavelength and flux calibration. Details of the reduction procedure are described in \cite{2026ApJ...999...10D}. 
The resulting spectra were then scaled using photometric data obtained at similar phases to place them on an absolute flux scale.  The spectra were corrected for the host-galaxy redshift ($z = 0.0075$). Extinction correction was then applied using the Milky Way reddening value of ${E(B - V )} = 0.02 \pm 0.0003$ mag, assuming negligible contribution from the host galaxy (see \S\ref{distance_extinction_explosion_epoch} for more details).

\section{Distance, extinction, and epoch of first light} 
\label{distance_extinction_explosion_epoch}

The redshift of the host galaxy was estimated from the heliocentric radial velocity measurement \cite{2000ApJ...529..786M}\footnote{\url{https://ned.ipac.caltech.edu/}}, yielding a value of $z = 0.0075 \pm 0.000003$. Owing to the proximity of the object, no additional correction for Virgo infall was applied.
 Assuming a cosmological model with $H_0 = 73 ;\mathrm{km};\mathrm{s}^{-1},\mathrm{Mpc}^{-1}$, $\Omega_{\Lambda} = 0.73$, and $\Omega_M = 0.27$, we derive a distance modulus of $\mu = 32.45 \pm 0.15$ mag, corresponding to a distance of $30.9 \pm 2.1$ Mpc.

 The Galactic reddening in the direction of the supernova was obtained as ${E(B - V )} = 0.02 \pm 0.0003$ mag \citep{2011ApJ...737..103S}. As no Na ID absorption feature at the host redshift was detected in the spectra, the extinction within the host galaxy is assumed to be negligible. 
Adopting a standard value of $R_V = 3.1$, this corresponds to a visual extinction of $A_V = 0.062 \pm 0.0009$ mag.

The epoch of first light for SN 2025qe was constrained using both observational limits and early-time light-curve modeling. The last non-detection reported on the Transient Name Server (TNS)\footnote{\url{https://www.wis-tns.org/object/2025qe}} is at JD 2460691.88, while the first detection occurred at JD 2460693.76. To further refine the explosion epoch, we modeled the early-time $g$ and $r$ band light curves using a power-law model of the form $F(t)=A(t-t_0)^n$, where $F(t)$ is the observed flux, $t_0$ is the time of first light, n is the rise index, and $A$ is a normalization constant. This model is motivated by the expanding fireball scenario, in which the early luminosity rises with the increasing surface area of the ejecta and is expected to follow $F(t) \propto t^2$ under idealized conditions of nearly constant temperature and expansion velocity \citep{1999AJ....118.2675R,2020ApJ...902...47M}. More generally, allowing $n$ to vary provides a flexible empirical description of the early rise and enables a constraint on the explosion epoch from pre-maximum photometry \citep{2022ApJ...925..217D}. The fit was restricted to the early rising phase ($\leq 50\%$ of the peak flux), where the power-law approximation is physically justified. Photometry from the Zwicky Transient Facility (ZTF) was combined with our pre-maximum observations to improve early-time coverage. 

From this model, we obtain an explosion epoch of $JD_{exp} = 2460692.44_{-0.44}^{+0.29}$, derived from the average of the independent $g$- and $r$-band fits. This estimate is in good agreement, within uncertainties, with the value reported by \cite{2025MNRAS.543.3731M}, who employed a more robust multi-band early-time fitting procedure using combined survey data. We therefore adopt this explosion epoch for subsequent analysis. The inferred rise index of $n \approx 1$ deviates from the canonical $n=2$ expected for a simple expanding fireball, indicating a shallower rise that may reflect mixing of $^{56}$Ni, indicating a thoroughly mixed ejecta similar to predictions from pure deflagration explosions, as also pointed out by \cite{2025MNRAS.543.3731M}. However, this estimate is subject to systematic uncertainties arising from the choice of fitting window, the assumption of a single power-law form, and the limited early-time sampling, and should therefore be interpreted with caution.

\section{Light curve and Color Curve}
\label{light_curve}

\subsection{Light curve Properties and Analysis}
\label{light_curve_color_curve}

\textit{Swift}/UVOT $U,B$ observations of SN 2025qe was combined with ground-based $u',g',r',i',z'$ observations from GIT and HCT to track its photometric evolution. 
 The HCT $u'$-band magnitudes were transformed to the Bessell $U$ system using the empirical relations of \cite{2005AJ....130..873J} to ensure a uniform photometric system. 
The resulting multi-band $U,B,g',r',i',z'$ light curves are shown in Figure~\ref{fig:SN_2025qe_light_curve}; the well-sampled photometric coverage in every filter allows us to determine the key light-curve parameters (peak magnitude $m_{\text{max}}$, epoch of maximum $\text{JD}_{\text{max}}$, and post-maximum decline rate $\Delta m_{15}$) in each filter. These parameters and their uncertainties were derived using the same method as described in \cite{2026ApJ...999...10D}, an approach which naturally takes into account both photometric noise and the limitations imposed by temporal sampling.

\begin{deluxetable*}{lcccccc}
\label{tab:lc_params}
\tablecaption{Multiband Light-curve Parameters of SN~2025qe}
\tablehead{
\colhead{Filter} & \colhead{$\lambda_{\rm eff}$ (\AA)} & \colhead{JD$_{\rm max}$} & \colhead{$m_{\rm max}$(mag)} & \colhead{$M_{\rm max}$(mag)} & \colhead{$\Delta m_{15}(\lambda)$(mag)} & \colhead{$t_R$(days)}
}
\startdata
$U$ & 3465 & $2460701.86 \pm 0.46$ & $16.25 \pm 0.06$ & $-16.30 \pm 0.16$ & $2.27 \pm 0.17$ & 9.42 \\
$g'$ & 4770 & $2460704.77 \pm 0.90$ & $16.36 \pm 0.03$ & $-16.16 \pm 0.15$ & $1.60 \pm 0.13$ & 12.33 \\
$r'$ & 6231 & $2460709.75 \pm 0.59$ & $16.29 \pm 0.06$ & $-16.21 \pm 0.16$ & $0.68 \pm 0.10$ & 17.31 \\
$i'$ & 7625 & $2460711.57 \pm 0.26$ & $16.40 \pm 0.02$ & $-16.09 \pm 0.15$ & $0.50 \pm 0.03$ & 19.13 \\
$z'$ & 9130 & $2460712.44 \pm 0.41$ & $16.49 \pm 0.03$ & $-15.99 \pm 0.15$ & $0.48 \pm 0.03$ & 20.00 \\
\enddata
\end{deluxetable*}

Table~\ref{tab:lc_params} lists the derived light curve parameters. The epoch of maximum light (${\text{JD}}_{\text{max}}$) shifts systematically to later times with increasing wavelength, and thus shows a gradually increasing rise time ($t_{R}$) from $U$ to $z'$ bands. the $U$-band peak occurs at JD $2460701.86\pm0.46$, followed by $g'$-band maximum $\sim$ 2.9 days later, and successively later maxima in $r'$, $i'$, and $z'$, with the reddest band peaking $\sim$ 10.6 days after the $U$-band peak. This blue-to-red lag ($U$ to $z'$) in attaining maxima is comparable to, though a little longer than, the lags reported for other well-observed Iax events -- for example, SN~2020kyg \citep[10.1~d,][]{2022MNRAS.511.2708S}, SN~2022eyw \citep[9.8~d,][]{2026ApJ...999...10D}, SN~2024pxl \citep[8.7~d,][]{2026ApJ...999..227S}, and reflects the progressive cooling of the receding photosphere, whose peak spectral energy distribution shifts redward following Wien's law as the ejecta expand and the effective temperature drops. In terms of luminosity, the $U$ band is the most luminous with a peak absolute magnitude of $M_U = -16.30 \pm 0.16$~mag. The peak $g'$-band magnitude of $M_{g'} = -16.16 \pm 0.15$~mag places it in between the brighter and fainter ends of the Iax class, with magnitudes similar to SN 2019muj ($M_g = -16.48 \pm 0.07$~mag; \citealt{2021MNRAS.501.1078B}) and SN 2024pxl ($M_g = -16.60 \pm 0.19$~mag; \citealt{2026ApJ...999..227S}) 

The post-maximum decline rate, $\Delta m_{15}(\lambda)$, shows pronounced wavelength dependence, falling from $\Delta m_{15}(U) = 2.27\pm0.17$ mag to $\Delta m_{15}(z') = 0.48\pm0.03$ mag. The steep decline of the $U$- and $g$-band light curves -- fading by more than two and $\sim$ 1.6 mag, respectively, within 15 days of maximum -- is driven by the onset of line blanketing from Fe II and Co II as the ejecta cools and recombines, which suppresses the emergent UV/blue flux and redistributes that energy towards longer wavelengths. The relatively shallow decline in $r', i'$ and especially $z'$ indicates that these bands continue to be powered by flux reprocessed from the blue, consistent with the trend seen across the SN~Iax class in which the redder filters retain their brightness longer as cooling proceeds \citep{2000ApJ...530..744P,2007ApJ...656..661K,2022NatAs...6..568N}. Unlike normal SNe Ia, no pronounced secondary maximum is seen in the redder bands. The absence of this feature in SNe~Iax is generally attributed to their lower total ejecta mass and to a more thoroughly mixed ejecta structure relative to normal Type~Ia explosions, which smoothens out the opacity change associated with Fe-group recombination that otherwise produces the secondary peak \citep{2006ApJ...649..939K,2022A&A...658A.179L}. The same suppressed or absent secondary maximum has been reported for other intermediate- and high-luminosity Iax events, including SN~2019muj \citep{2021MNRAS.501.1078B}, SN~2024pxl \citep{2026ApJ...999..227S}, SN~2020kyg \citep{2022MNRAS.511.2708S} and SN~2022eyw \citep{2026ApJ...999...10D}.

In absolute terms, SN~2025qe is fainter and evolves in the blue bands more rapidly than the luminous Iax events such as SN~2005hk \citep{2007PASP..119..360P}, SN~2012Z \citep{2015A&A...573A...2S}, and SN~2022eyw both of which peak near $M_{B/g}\sim-18$ mag and decline more gradually ($\Delta m_{15}(B/g)\lesssim1.5$ mag). At the same time, SN~2025qe is considerably more luminous and slower-declining than the low-luminosity events SN~2008ha \citep{2009AJ....138..376F}, SN~2010ae \citep{2014A&A...561A.146S}. This combination of intermediate peak luminosity, rapid blue-band fading, but comparatively broad red/near-IR light curves places SN~2025qe in the luminosity gap between the bright and faint extremes of the Iax class -- the same transitional territory occupied by the well-studied intermediate-luminosity events SN~2019muj and SN~2024pxl. The small values of $\Delta m_{15}(r')$, $\Delta m_{15}(i')$, and particularly $\Delta m_{15}(z')$ argue for a relatively long photon diffusion time in the inner ejecta, broadly consistent with the intermediate ejecta masses inferred for SN~2019muj and SN~2024pxl from their bolometric light-curve modeling. A detailed discussion on this is given in Section \ref{fig:bolometry_comparison}.

\begin{figure*}
	\begin{center}
		 {\includegraphics[width=0.8\linewidth]{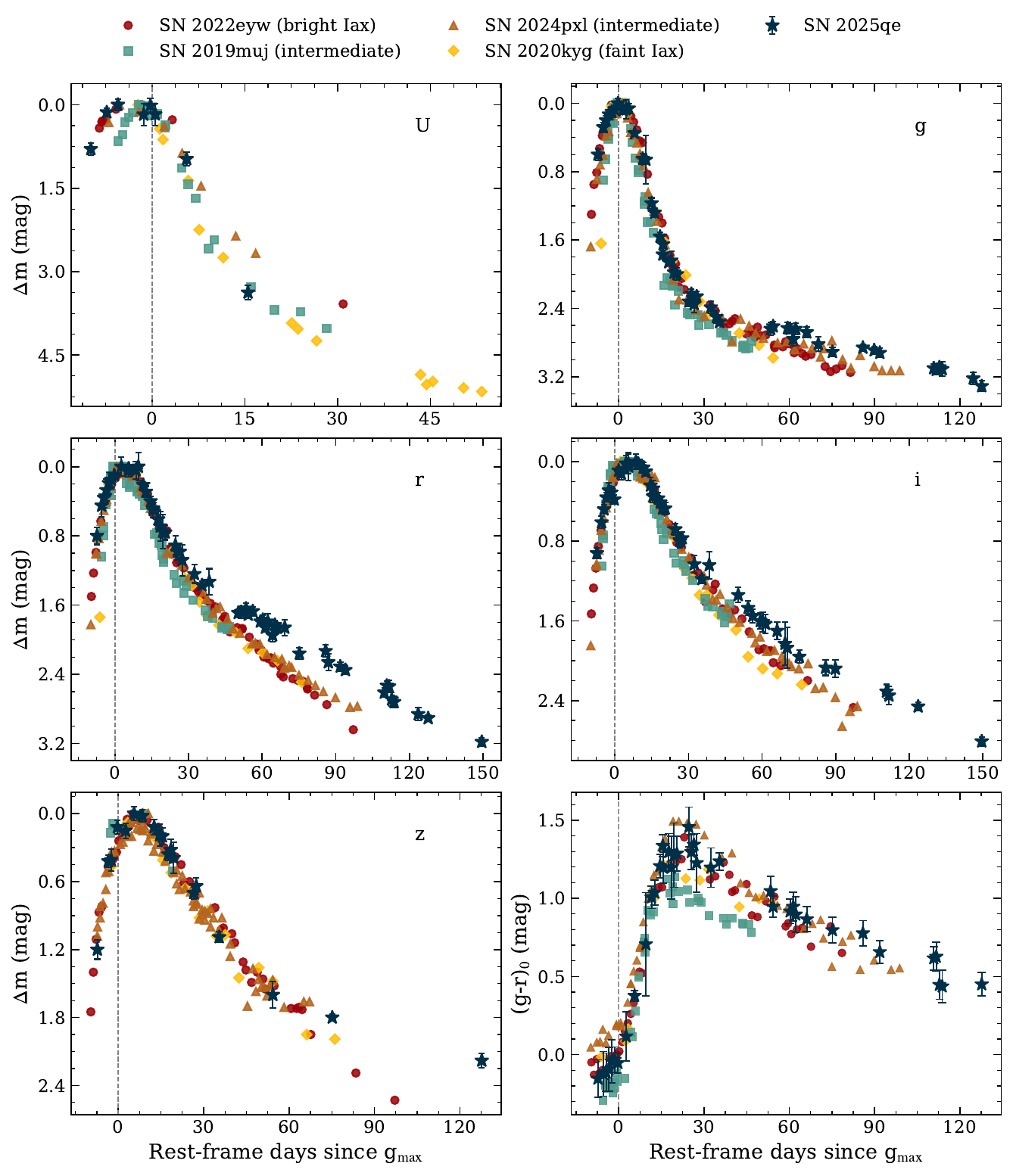}}
	\end{center}
	\caption{Normalized $Ugriz$-band light curves of SN~2025qe compared with those of other Type~Iax supernovae. The light curves are normalized to their respective peak magnitudes to facilitate comparison of their post-maximum evolution. The vertical grey dashed line in each panel marks the epoch of $g$-band maximum. The progressively broader evolution of SN~2025qe towards longer wavelengths is evident, particularly in the $r$, $i$, and $z$ bands. The lower-right panel shows the comparison of dereddened $(g-r)_0$ colour evolution of SN~2025qe with other Iax SNe, illustrating the corresponding evolution of their spectral energy distributions.}
	\label{fig:lc_gr_comparison}
\end{figure*}

\begin{figure}
	\begin{center}
		 \resizebox{\hsize}{!}{\includegraphics{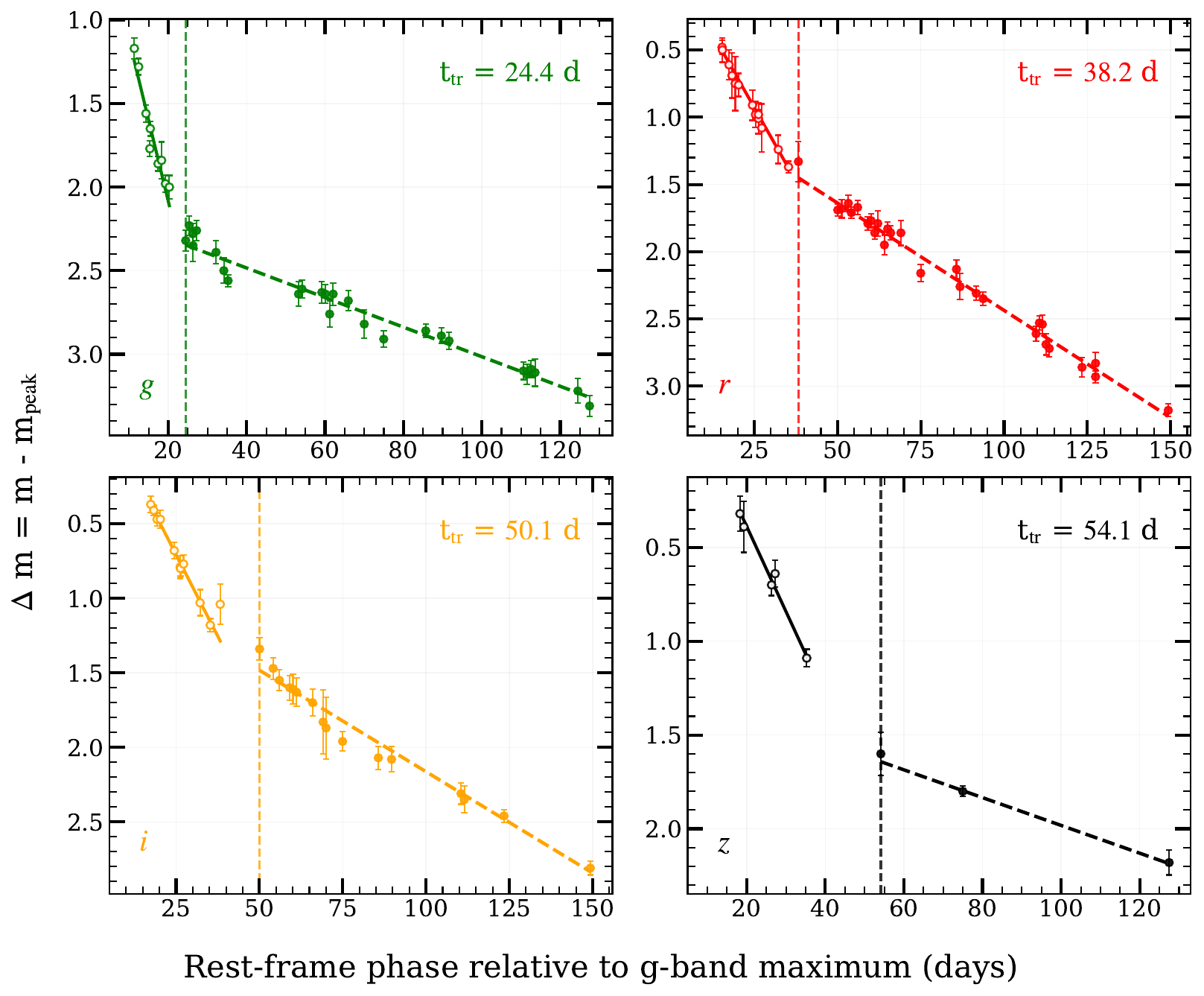}}
	\end{center}
	\caption{The wavelength-dependent onset of light-curve broadening in SN~2025qe, quantified using a broken-line fit to the post-maximum light curves. The vertical dashed lines mark the adopted transition epochs, $t_{\rm tr}$, at which the light curves transition to a slower-declining phase. 
    The solid and dashed lines show the weighted linear fits to the early and late branches, respectively. The light curves are normalized to their respective peak magnitudes and plotted relative to the $g$-band maximum.}
	\label{fig:lc_fit}
\end{figure}

\subsection{Light Curve and Color Curve Comparison}
\label{lc_comparison}

Figure \ref{fig:lc_gr_comparison} shows the comparison of the normalized $Ug'r'i'z'$ light curves of SN 2025qe with those of several well-observed SNe Iax for which observations in the same filters were available. The photometry of SN~2025qe was obtained in the SDSS primed $g'r'i'z'$ filters, whereas the comparison SNe were primarily observed in the corresponding unprimed $g,r,i,z$ bands. Since the differences between the filter response functions and effective wavelengths are minimal, we will refer to them as $griz$ from now on for consistency. SDSS $u'$ to Bessell $U$ band conversion has been done wherever necessary using the color transformations of \cite{2005AJ....130..873J}. The light curves have been shifted such that the peak magnitude in each band corresponds to $\Delta m = 0$, enabling a direct comparison of their post-maximum evolution. In the $U$ band, SN 2025qe exhibits an evolution that is largely consistent with the comparison sample. A mild deviation becomes apparent in the $g$ band, where SN 2025qe declines slightly more slowly at intermediate and late phases. However, the increased width of the light curves becomes significantly more pronounced in the redder bands, particularly from the $r$ band onward. The divergence from the comparison objects first appears in the $g$ band and subsequently develops in the $r$, $i$, and $z$ bands at progressively later epochs. As a result, SN 2025qe remains systematically brighter than the comparison SNe Iax at late times, with the effect becoming stronger toward longer wavelengths. This wavelength-dependent brightening suggests an evolving spectral energy distribution, consistent with a gradual redistribution of flux toward redder wavelengths as the supernova evolves. This shallow decline at progressively redder wavelengths, which is attributed to the increasing influence of Fe-group line opacity and fluorescence \citep{2009MNRAS.398.1809K}, may point toward a higher Fe-group-enriched ejecta in case of SN 2025qe than in the other Iax. 

The $(g - r)$ color evolution indicates that SN~2025qe follows the general behavior observed in other SNe Iax (See lower right panel of Fig. \ref{fig:lc_gr_comparison}). Prior to maximum light, the object exhibits relatively blue colors, followed by a rapid reddening phase after peak brightness. The $(g-r)$ color reaches a maximum of $\sim$ 1.4 mag around 20 - 30 days after $g$ band maximum and subsequently evolves toward bluer colors at later epochs. Although SN~2025qe exhibits progressively broader light curves toward longer wavelengths, its $(g - r)$ color evolution closely resembles that of other SNe Iax. Throughout its evolution, the color curve of SN 2025qe remains broadly consistent with those of SN 2022eyw, SN 2019muj, SN 2024pxl, and SN 2020kyg, showing no evidence for unusually red colors despite its comparatively broader red-band light curves. This suggests that the enhanced late-time emission observed in the redder bands of SN 2025qe is not accompanied by a significant deviation in its overall optical color evolution, and instead may arise from a more gradual redistribution of flux across the optical wavelength range.

 To quantify the epoch at which the normalized optical light curves of SN 2025qe start to become progressively broader at longer wavelengths relative to other SNe Iax, we modeled the post-maximum evolution in each band using a broken-line fit (Figure \ref{fig:lc_fit}).

For each filter, the light curve was first normalized with respect to the peak magnitude and then the data obtained more than 10 days after maximum were considered in the fit in order to exclude the rapidly evolving peak phase. 
The transition epoch was determined through a grid-search procedure. For a set of candidate break times, the light curve was divided into two segments and each segment was independently fit with a weighted linear relation using the photometric uncertainties as weights. The total goodness-of-fit statistic,

\begin{equation}
    \chi^2_{\rm tot}=\chi^2_1+\chi^2_2,
\end{equation}

was computed for each candidate break, and the optimal break time was taken to be the value that minimized $\chi^2_{\rm tot}$. We define the transition epoch, ($t_{\rm tr}$), as the onset of the shallower decline branch identified by the broken-line fit. This time is expressed as the phases relative to the $g$-band maximum in order to enable a direct comparison among different filters.

The resulting transition epochs are $t_{\rm tr}(g)\approx24.4\ {\rm d}$,
$t_{\rm tr}(r)\approx38.2\ {\rm d}$, $t_{\rm tr}(i)\approx50.1\ {\rm d}$, and $t_{\rm tr}(z)\approx54.1\ {\rm d}$ measured relative to the epoch of $g$-band maximum.
Following the transition, the decline rates become substantially shallower. The post-transition decline rates measured from the late-time linear component are $0.88 \pm 0.04\ {\rm mag\ 100\;d^{-1}}$ in $g$, $1.60 \pm 0.03\ {\rm mag\ 100
;d^{-1}}$ in $r$, $1.36 \pm 0.05\ {\rm mag\ 100\;d^{-1}}$ in $i$, and $0.74 \pm 0.05\ {\rm mag\ 100\;d^{-1}}$ in $z$. These values are significantly lower than the corresponding early-time decline rates, confirming the presence of a distinct flattening phase in the optical evolution of SN 2025qe.

\subsection{Bolometric Light Curve Analysis}
\label{pseudo-bolometry}

\begin{figure}
	\begin{center}
		{\includegraphics[width=0.5\textwidth]{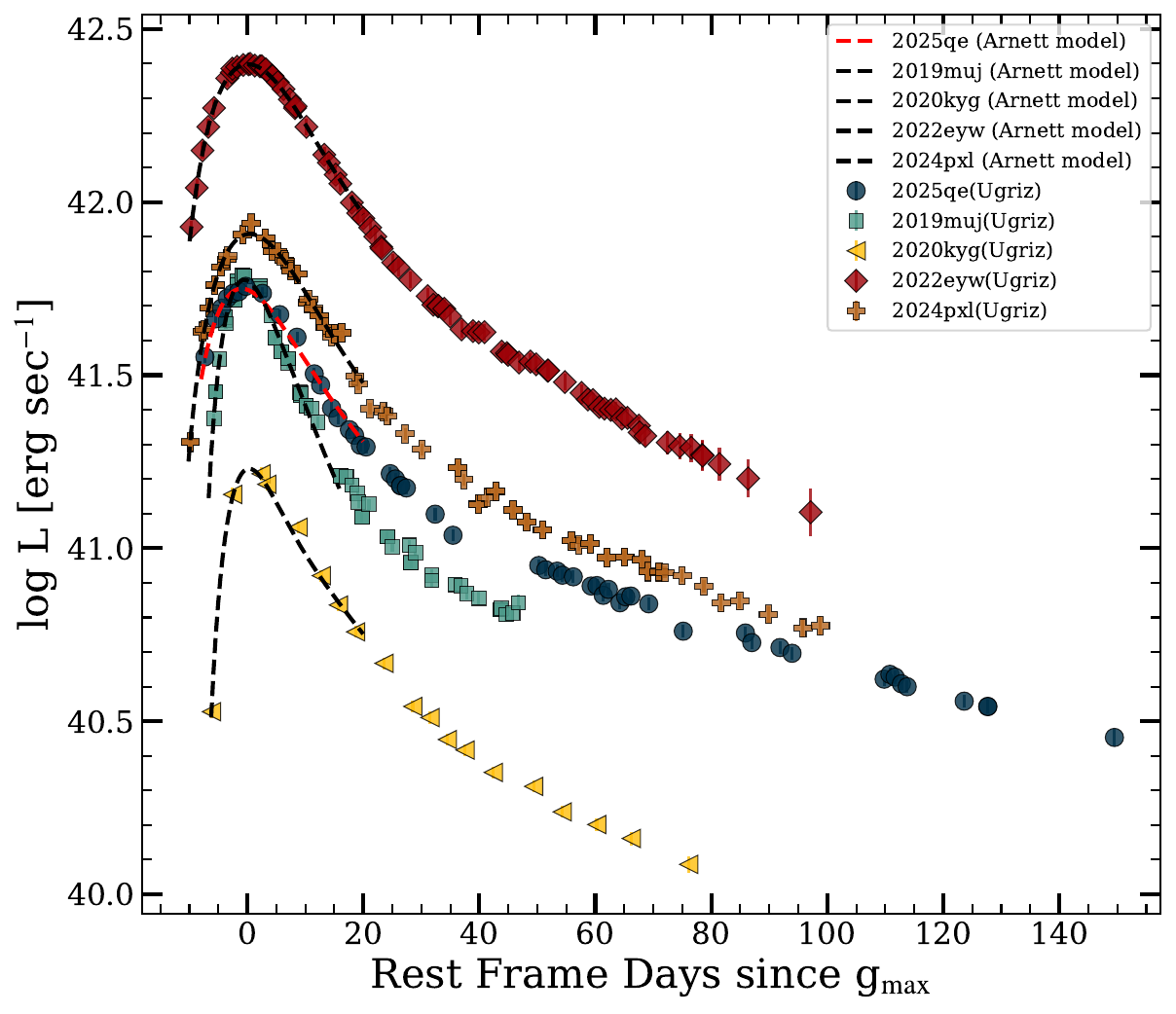}}
	\end{center}
	\caption{Comparison of the \(Ugriz\) pseudobolometric light curve of SN 2025qe with those of other SNe Iax SN 2019muj, SN 2020kyg, SN 2022eyw, and SN 2024pxl. The red dashed curve shows the best-fit Arnett model for SN 2025qe, while the black dashed curves show the corresponding Arnett fits for the comparison SNe. For consistency, the Arnett models are shown only up to \(+20\) rest-frame days since the \(g\)-band maximum, corresponding to the early-time phase adopted for the Arnett modeling.}
	\label{fig:bolometry_comparison}
\end{figure}

The pseudobolometric light curve of SN~2025qe was obtained with the \textsc{SuperBol} code \citep{2018RNAAS...2..230N} from the extinction-corrected $U,g,r,i,z$ photometry. The dereddened magnitudes were converted to flux densities using the standard zero-points and effective wavelengths of each filter, and the resulting spectral energy distribution was integrated at each epoch using the trapezoidal rule to obtain the pseudobolometric luminosity. The relatively good temporal sampling allowed us to construct the pseudobolometric light curve directly from the observed data, with uncertainties propagated from the photometric errors.

The photometric coverage of SN~2025qe is sufficiently dense in the $g$, $r$, $i$, and $z$ bands, and therefore no interpolation between filters was required for constructing the optical spectral energy distribution. However, the $U$-band observations extend only up to JD 2460720, after which its contribution is no longer directly available.

To estimate the missing $U$-band contribution at later phases, we used SN~2019muj \citep{2021MNRAS.501.1078B} as a reference object. SN~2019muj is an intermediate-luminosity Type Iax supernova with well-sampled $Ugriz$ photometry and exhibits photometric properties comparable to those of SN~2025qe. In particular, the normalized $U$-band light curves of the two objects show very similar evolution (see Section \ref{lc_comparison}), making SN~2019muj a suitable empirical template for estimating the missing flux.

We first reconstructed the $Ugriz$ and $griz$ pseudobolometric light curves of SN~2019muj using the same procedure. From these two light curves, we determined the fractional contribution of the $U$ band to the total pseudobolometric luminosity as a function of phase. After the last available $U$-band observation of SN~2025qe, this contribution becomes small and evolves only slowly with time. We therefore used the corresponding fractional contribution derived from SN~2019muj to estimate the missing $U$-band flux ($\sim 12\%$) and added it to the $griz$ pseudobolometric luminosity of SN~2025qe.

This correction should only be regarded as a rough approximation, since SNe Iax exhibit intrinsic diversity in their photometric evolution. Nevertheless, the adopted correction has only a modest effect on the late-time luminosity because the $U$-band contribution is already small at later epochs \citep{2017hsn..book..375J}. Moreover, the uncertainty introduced by the adopted late-time $U$-band correction does not significantly affect the derived $^{56}$Ni mass or other explosion parameters estimated using the Arnett semi-analytical modeling presented in Section~\ref{Arnett_Modeling}. This is because the Arnett model is constrained primarily by the early-phase bolometric light curve, when the $U$-band observations are available, and the assumptions of homologous expansion and radiative diffusion through an approximately spherical ejecta are expected to hold \citep{1982ApJ...253..785A,2008MNRAS.383.1485V}.

\subsubsection{Analytical Estimate of Radioactive \texorpdfstring{$^{56}$}{}Ni and other Explosion Parameters}
\label{Arnett_Modeling}

\noindent The amount of radioactive $^{56}$Ni produced in the explosion was estimated from the pseudo-bolometric light curve shown in Figure~\ref{fig:bolometry_comparison}. 

We fit the semi-analytical Arnett luminosity model to the $Ugriz$ pseudo-bolometric light curve of SN 2025qe to estimate its explosion and ejecta properties. The fitting was restricted to the early photospheric evolution, where the assumptions underlying the diffusion approximation remain valid. The free parameters of the model were the explosion epoch ($t_{\rm exp}$), the effective photon diffusion timescale ($t_{\rm lc}$), the gamma-ray leakage timescale ($t_{\gamma}$), and the synthesized $^{56}$Ni mass ($M_{\rm Ni}$). The parameter space was explored using a Markov Chain Monte Carlo (MCMC) approach implemented with the Python package \texttt{emcee}, which samples the posterior probability distributions of the model parameters. The median values of the posterior distributions are adopted as the best-fit estimates, while the quoted uncertainties correspond to the 16th and 84th percentiles. A detailed description of the semi-analytical luminosity formalism is provided in \cite{2026ApJ...999...10D}. The resulting best-fitting model is shown in Figure \ref{fig:bolometry_comparison}. The MCMC analysis yields an explosion epoch of $JD_{\rm exp}=2460690.54_{-0.34}^{+0.41}$, a synthesized $^{56}$Ni mass of $0.02_{-0.0005}^{+0.0005}\;M_\odot$, a $t_{\rm lc}$ of $10.69_{-0.49}^{+0.40}$ d, and a $t_{\gamma}$ of $44.40_{-1.04}^{+1.11}$ d.

The ejecta mass $M_{\rm ej}$ and kinetic energy $E_{\rm k}$ were derived from the diffusion model parameters following the canonical relations of (\citealt{1982ApJ...253..785A}):

\begin{equation}
\label{eq:Mej}
M_{\rm ej} = 0.5 \frac{ \beta c}{ \kappa}v_{\rm ph} t_{lc}^2,
\end{equation}

\begin{equation}
\label{eq:Ek}
E_{\rm k} = 0.3 M_{\rm ej} v_{\rm exp}^2,
\end{equation}

\noindent where $\beta = 13.8$ is the dimensionless integration constant from the diffusion solution, $\kappa_{\rm opt} = 0.1$ cm$^2$ g$^{-1}$ represents the characteristic opacity of Fe-group elements in the ejecta \citep{2000ApJ...530..757P,2015MNRAS.453.2103S,2020ApJ...892L..24S}, $c$ is the speed of light, and $v_{\rm exp} \approx 5100$ km s$^{-1}$ is the ejecta expansion velocity inferred from the Si~II $\lambda$6355 absorption minimum near $g$-band maximum \citep{2025MNRAS.543.3731M}. Substituting these values into Equations~(\ref{eq:Mej}) and (\ref{eq:Ek}) yields an ejecta mass of $M_{\rm ej}=0.45^{+0.035}_{-0.042},M\odot$ and a kinetic energy of $E_{\rm k}=7.02^{+0.53}_{-0.64}\times10^{49}$ erg for SN~2025qe.

\subsubsection{Comparison and Analysis of the Pseudo-Bolometric Light Curve}
\label{bolometry_comparison}

The $Ugriz$ pseudobolometric light curve of SN~2025qe was compared with those of other SNe Iax whose observations were available in the same bands. Since SNe Iax span a wide range in peak brightness, decline rate, and radiated energy, such a comparative study is essential for understanding where this event lies in comparison to other SNe Iax in terms of peak luminosity, Ni mass, ejecta mass, kinetic energy. 

As in other SNe Iax, the luminosity evolution is governed primarily by the diffusion of photons produced by the decay chain $^{56}\mathrm{Ni} \rightarrow ^{56}\mathrm{Co} \rightarrow ^{56}\mathrm{Fe}$,
while the detailed shape of the light curve is regulated by the ejecta mass, kinetic energy and opacity \citep{2000ApJ...530..744P,2000ApJ...530..757P}. A larger ejecta mass increases the diffusion time, producing broader light curves \citep{2018MNRAS.474.3187W}. Figure \ref{fig:bolometry_comparison} shows how the pseudobolometric light curve of SN~2025qe evolves in comparison to other Iax events. The Arnett semi-analytical model fits to the $Ugriz$ pseudo-bolometric light curves of all the SNe are also shown. The corrected bolometric curve of SN~2025qe places it in the intermediate-luminosity range occupied by objects like SN~2019muj \citep{2021MNRAS.501.1078B} and SN~2024pxl \citep{2026ApJ...999..227S}. For a robust comparative analysis, the explosion parameters of these objects have been compared alongside their peak luminosities (See Table \ref{tab:bolometry_comparison}). The peak luminosity of SN 2025qe is similar to that of SN~2019muj, but the diffusion time is markedly longer, leading to a wider light curve. This longer diffusion time may be associated with the larger ejecta mass inferred for SN~2025qe compared with SN~2019muj. At the same time, the ejecta mass, photon diffusion time, and hence the light-curve width of SN~2025qe closely track those of SN~2024pxl. The kinetic energy ($E_{\text{k}}$), however, is substantially higher in SN~2025qe than both SN~2019muj and SN~2024pxl, owing to its higher expansion velocity ($\sim5100$, $\sim4430$, and $\sim3800$ km s$^{-1}$, respectively; \citealt{2025MNRAS.543.3731M, 2021PASJ...73.1295K,2026ApJ...999..227S}) estimated from the Si~II $\lambda6355$ line velocity near maximum light. Taken together, SN~2025qe appears to occupy a transitional regime between these two intermediate-luminosity events, strengthening the interpretation that SNe Iax form a continuous family of explosions rather than a sharply separated set of subgroups \citep{2013ApJ...767...57F, 2017hsn..book..375J}.

From a physical point of view, the bolometric light curve suggests that SN~2025qe was powered by a moderate amount of radioactive $^{56}$Ni -- comparable to that inferred for SN~2019muj-- while its light-curve width and decline rate point to an ejecta mass and Ni distribution similar to that of SN~2024pxl. The overall evolution is compatible with a low-energy thermonuclear event involving incomplete burning and partial disruption of a WD, consistent with a pure (or weak) deflagration of a Chandrasekhar-mass CO WD that fails to fully unbind the star and may leave behind a bound remnant \citep{2007PASP..119..360P, 2012ApJ...761L..23J, 2013MNRAS.429.2287K, 2014MNRAS.438.1762F}. A detailed comparison of SN 2025qe with deflagration models is done in the following section.

\input{Bolometry_Comparison_Table}

\subsubsection{Comparison with Deflagration Models}
\label{sec:deflagration_comparison}

\begin{figure*}
    \centering
    \resizebox{\hsize}{!}{\includegraphics{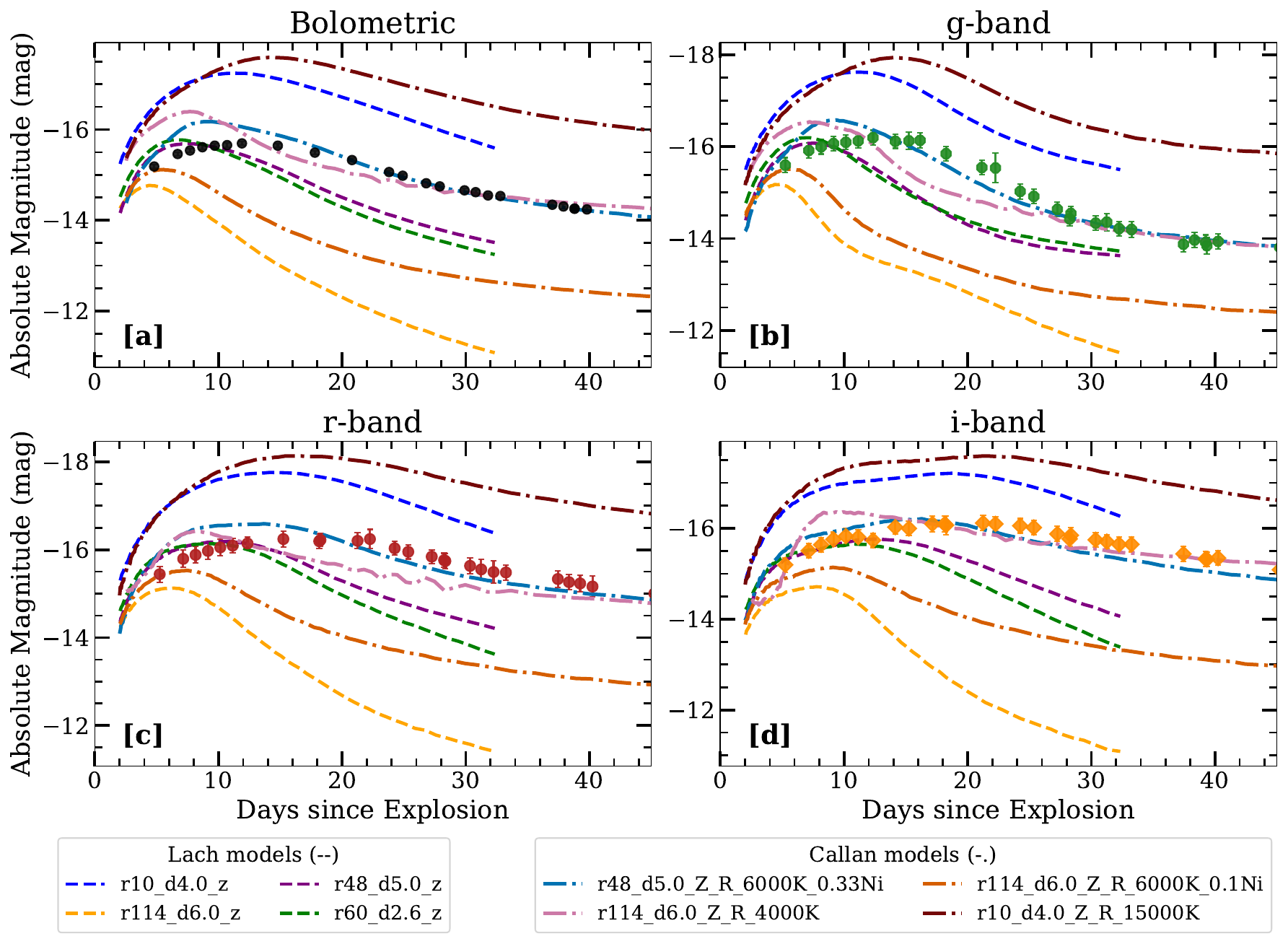}}
    \caption{$Ugriz$ pseudobolometric and multi-band $(gri)$ light-curve comparison of SN 2025qe with the synthetic light curves of selected three-dimensional pure deflagration models from \cite{2022A&A...658A.179L} and \cite{2024MNRAS.530.1457C}}
    \label{fig:deflagration_callan}
\end{figure*}

To investigate whether the photometric evolution of SN 2025qe can be explained by pure deflagration explosions, we compared its pseudobolometric and multiband light curves with the recent grid of three-dimensional hydrodynamic models presented by \citealt{2022A&A...658A.179L} (L22). Unlike the earlier deflagration calculations of \citealt{2014MNRAS.438.1762F} (F14), which primarily vary the ignition configuration, the L22 models explore a broader parameter space, which includes the ignition offset (`r'), central density (`d'), metallicity (`Z'), and rotational state of the progenitor WD. This approach allows the influence of individual progenitor and explosion properties on the observable characteristics of SNe Iax to be examined in a more systematic manner. Their study showed that the ignition offset and central density are the principal parameters governing the explosion strength and the amount of synthesized $^{56}$Ni, whereas metallicity and rotation play a comparatively minor role.

For the comparison presented here, pseudobolometric light curves were constructed from the synthetic spectral time series provided by \citealt{2022A&A...658A.179L}. The model spectra were integrated over the wavelength interval 3200-9967 \AA, corresponding to the maximum spectral coverage available in the published synthetic spectra. This wavelength range closely matches that used in constructing the observed $Ugriz$ pseudobolometric light curve of SN 2025qe. Although the observed $z$-band extends slightly beyond the upper integration limit of the models, the contribution from the interval 10000–11000 \AA~ is small, and thus does not significantly affect the resulting comparison.

Panel [a] of Figure \ref{fig:deflagration_callan} compares the pseudobolometric light curve of SN 2025qe with several representative L22 models. The luminosity peak of SN 2025qe matches closely with the r48\_d5.0\_Z and r60\_d2.6\_Z models. This agreement near maximum is not unexpected: both models synthesize approximately 0.02 $M_{\odot}$ of $^{56}$Ni, comparable to the $^{56}$Ni inferred for SN 2025qe from the Arnett-model analysis, and the peak luminosity of thermonuclear supernovae is primarily set by the radioactive energy released by $^{56}$Ni and its daughter isotope $^{56}$Co.

Despite this agreement at peak, the post-maximum evolution reveals a clear discrepancy: all standard L22 models decline significantly faster than the observed pseudobolometric light curve of SN 2025qe, with the difference becoming increasingly pronounced at later epochs. This indicates that the ejecta in the models become transparent to radioactive energy deposition earlier than required by the observations, a behaviour also reported for other intermediate-luminosity SNe Iax (e.g. \citealt{2016A&A...589A..89M}; \citealt{2022A&A...658A.179L}). The most plausible explanation is a difference in ejecta mass: although the models reproduce the nickel yield inferred for SN 2025qe, their ejecta masses remain substantially lower than the value derived from our light-curve modeling. A larger ejecta mass increases the diffusion timescale and improves $\gamma$-ray trapping, thereby sustaining the luminosity for longer after maximum. The same trend is seen in the multiband comparison of Figure \ref{fig:deflagration_callan}: while r60\_d2.6\_Z gives the closest match to the observed $g$-band peak, the discrepancy in decline rate grows systematically toward redder bands, with the model light curves becoming increasingly narrower relative to SN 2025qe from $r$ through $z$. This points to a larger fraction of the radiative output in SN 2025qe being redistributed to redder wavelengths at late times than the standard deflagration models predict -- again consistent with more efficient trapping and reprocessing of radioactive energy in a more massive ejecta.

Since the combined bolometric and multiband comparisons remain unable to account for the slow post-maximum evolution and enhanced red-band emission of SN 2025qe, we tested the effect of adding a bound remnant to the model light curves. For a wide range of ignition configurations, pure deflagrations of Chandrasekhar-mass WDs fail to fully unbind the star, leaving a remnant of $\gtrsim 1$ M$_{\odot}$ polluted with burning products, including a non-negligible fraction (2-7 per cent) of the synthesized $^{56}$Ni  \citep{2015MNRAS.450.3045K, 2022A&A...658A.179L}. Since this remnant material is not part of the homologously expanding ejecta, it is excluded by construction from the standard radiative transfer calculations, even though its radioactive decay continues to release energy that can, in principle, diffuse outward and contribute to the observed light curve at post-maximum epochs. \citealt{2024MNRAS.530.1457C} (C24) incorporated this effect directly into 3D time-dependent Monte Carlo radiative transfer simulations (\texttt{ARTIS}) built on the same \citealt{2022A&A...658A.179L} explosion models, by placing a central $^{56}$Ni-powered source at the remnant location and treating it as a blackbody with the injected radiation packets propagated self-consistently through the ejecta.

Our comparison to the modified C24 models is shown alongside the standard L22 models in Figure \ref{fig:deflagration_callan}. The inclusion of the bound remnant systematically broadens the light curves and slows the post-maximum decline relative to the corresponding standard model, with the effect most pronounced in the redder optical bands $(r,i)$ and comparatively modest in $g$, in line with what \citealt{2024MNRAS.530.1457C} find for their comparisons with bright, intermediate, and faint SNe Iax. In our comparison, the models r48\_d5.0\_Z\_R\_6000K\_0.33Ni and r114\_d6.0\_Z\_R\_4000K -- which adopt reduced remnant $^{56}$Ni masses of one-third and the full amount, respectively, at remnant temperatures of 6000 and 4000 K respectively -- track the observed decline of SN 2025qe considerably better than any of the unmodified Lach models, both in the pseudobolometric light curve and in the $r$ and
$i$-band light curves, where the standard models are most discrepant. The hotter, high-$^{56}$Ni remnant model (r10\_d4.0\_Z\_R\_15000K) overpredicts the luminosity at all post-peak epochs, while the strongly nickel-suppressed r114\_d6.0\_Z\_R\_6000K\_0.1Ni model falls below the observations, bracketing the range of remnant properties relevant for an intermediate-luminosity event such as SN 2025qe.

\section{Spectral Analysis}
\label{spectral analysis}

\subsection{Spectral Evolution}
\label{spectral_evolution}

\begin{figure*}
	\begin{center}
	   \includegraphics[width=0.75\linewidth]{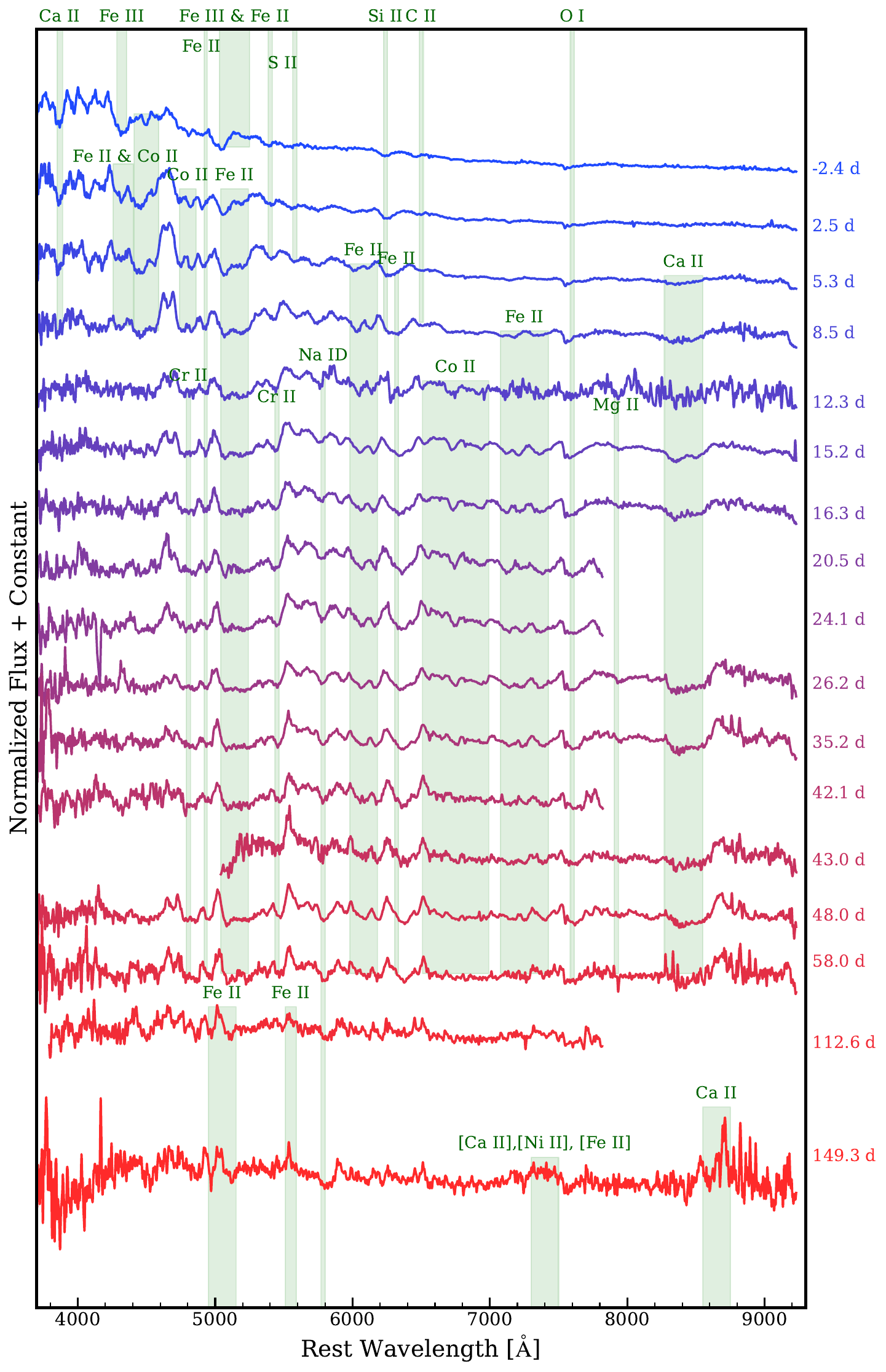}
	\end{center}
	\caption{Spectral evolution of SN 2025qe spanning from -2.4 to +149.3 rest-frame days relative to $g$-band maximum. All spectra have been corrected for reddening and redshift, and smoothened to enhance clarity. Line identifications in the first four spectra are primarily based on the TARDIS modeling (see Section~\ref{spectral_modeling}), while those in the later spectra are adopted from \cite{2008ApJ...680..580S,2013ApJ...767...57F,2021MNRAS.501.1078B,2022ApJ...925..217D,2025ApJ...988..209H}. The spectra are normalized by their median flux values.}
	\label{fig:spectra}
\end{figure*}

The spectral evolution of SN~2025qe from $-2.4$ to $+149.3$ rest-frame days relative to $g$-band maximum is shown in Figure \ref{fig:spectra}. Hereafter, all phases are quoted in rest-frame days relative to the $g$-band maximum, unless otherwise stated. The earliest spectrum exhibits a relatively blue continuum together with prominent absorption features of Fe\,{\sc iii}, indicating a moderately hot photosphere. Weak absorptions due to S\,{\sc ii} $\lambda\lambda$5454,5640, Si\,{\sc ii} $\lambda$6355, and C\,{\sc ii} $\lambda$6580 are also present. This simultaneous presence of Fe lines, together with discernible Si\,{\sc ii} and C\,{\sc ii} features in the pre-maximum spectrum, is commonly observed in SNe~Iax, particularly in intermediate- and faint-luminosity events \citep{2022MNRAS.517.5617S, 2023ApJ...953...93S}, and suggests incomplete burning in the outer ejecta, consistent with predictions from pure-deflagration explosion models \citep{2025ApJ...988..209H, 2022A&A...658A.179L}. As the supernova evolves toward maximum light, the Fe\,{\sc iii} features near 4300 and 5000~\AA\ get replaced by Fe\,{\sc ii}, indicating the recombination of the ejecta as the photosphere cools. Some Co\,{\sc ii} lines also start appearing, mixed with the Fe\,{\sc ii} lines between 4300 and 4600~\AA. The S\,{\sc ii} ``W" feature remains identifiable through the earliest spectra but fades rapidly after maximum light. The Si\,{\sc ii} 6355 absorption remains relatively weak throughout the photospheric phase and becomes difficult to identify beyond $\sim$+10 d. The C\,{\sc ii} 6580 feature is visible in the earliest spectra but disappears shortly after maximum light, indicating that the outer carbon-rich layers are no longer contributing significantly to the observed spectrum. 

The most notable change in the post-maximum evolution is the increasing dominance of iron-group elements. By $\sim$ 10–15 d, Fe\,{\sc ii} becomes the principal contributor to the optical spectrum, producing broad absorption complexes around 5000, 6100, and 7300 \AA. At the same time, Cr\,{\sc ii} features near 4800 and 5500 \AA emerge and strengthen, while broad Co\,{\sc ii} absorption develops between approximately 6500 and 7000 \AA. The appearance of these lower-ionization species is consistent with the progressive cooling and recombination of the ejecta. Similar post-maximum evolution, characterized by increasing contributions from Fe\,{\sc ii}, Cr\,{\sc ii}, and Co\,{\sc ii}, is observed across the luminosity sequence of SNe~Iax, although the relative strengths and timescales of these features vary among individual events \citep{2008ApJ...680..580S, 2014A&A...561A.146S, 2021MNRAS.501.1078B, 2026ApJ...999..227S}.

The red portion of the spectrum also undergoes a significant evolution. The O\,{\sc ii} 7774 and Mg\,{\sc ii} 7877,7896 features become increasingly evident during the first few weeks after maximum, while the Ca\,{\sc ii} near-infrared triplet strengthens steadily and remains one of the most prominent features throughout the post-maximum evolution. Between $+$20 and $+$60 d, the spectra evolve relatively slowly and are dominated by broad, blended absorption features of Fe\,{\sc ii}, Cr\,{\sc ii}, Co\,{\sc ii}, and Ca\,{\sc ii}. The persistence of these permitted lines and the absence of a fully nebular spectrum indicate that the ejecta remain at relatively high densities during this phase.

By +112.6 d, the spectrum is still dominated by permitted Fe\,{\sc ii} features, demonstrating the characteristically slow spectroscopic evolution of the Type Iax class. At the last epoch ($+$149.3 d), weak forbidden emission begins to emerge in the vicinity of 7300 \AA, where a blend of [Ca\,{\sc ii}], [Ni\,{\sc ii}], and [Fe\,{\sc ii}] is present. Emission associated with the Ca\,{\sc ii} near-infrared triplet also remains prominent. Nevertheless, the continued presence of strong permitted Fe\,{\sc ii} lines indicates that SN~2025qe has not yet fully transitioned into the nebular phase. Such delayed nebular evolution is a defining property of SNe Iax and is generally attributed to their low ejecta velocities, high inner ejecta densities, and the possible presence of a bound remnant following a failed or partial deflagration \citep{2006AJ....132..189J,2016MNRAS.461..433F}.

\begin{figure}
    \centering
    { \includegraphics[width=\linewidth]{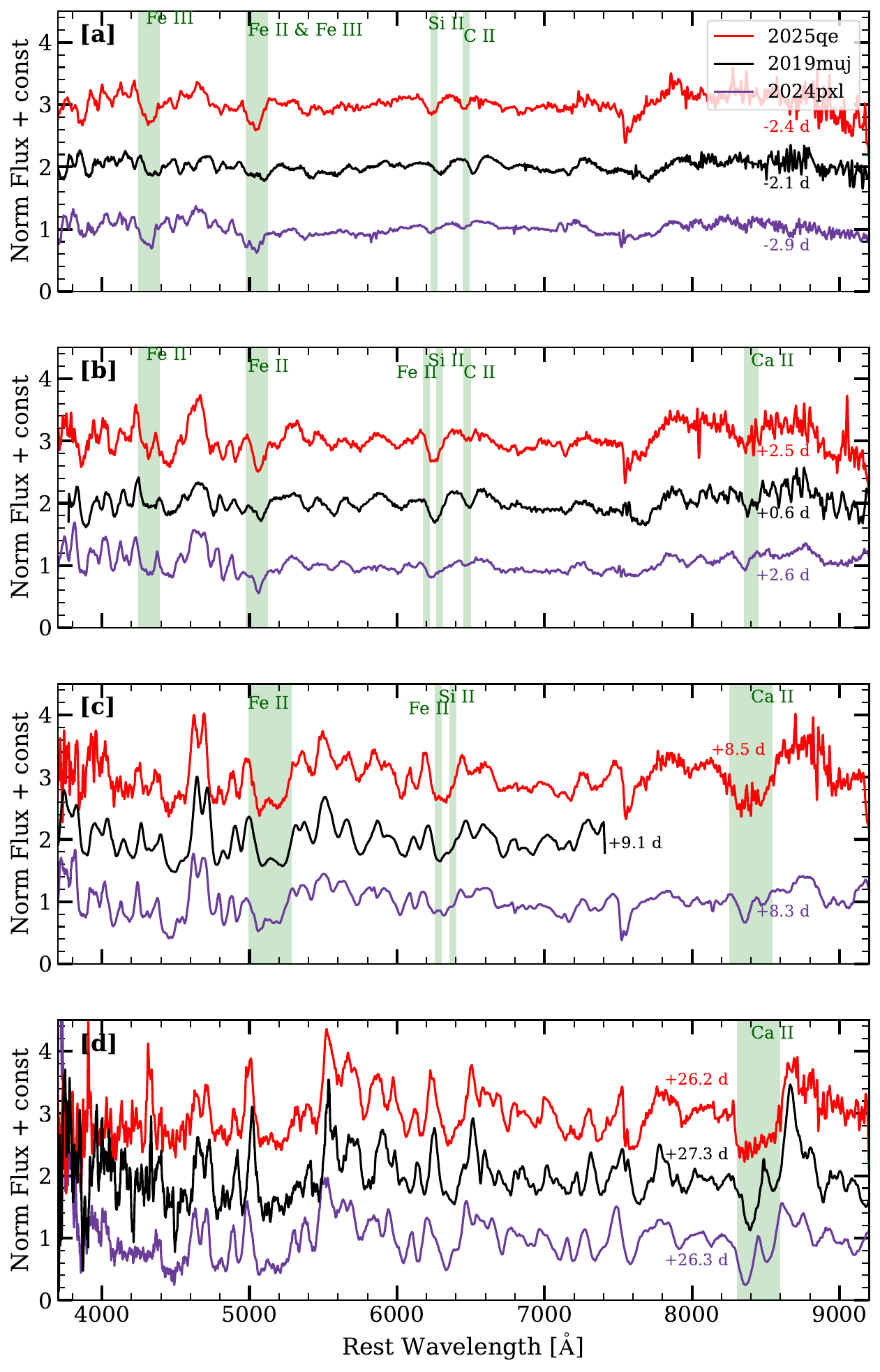}}
    \caption{Comparison of SN 2025qe spectra with those of the spectroscopically similar, intermediate-luminosity SNe Iax SN 2019muj and SN 2024pxl at four representative epochs. The spectra are vertically offset for clarity and continuum-normalized to facilitate comparison of the spectral features.}
    \label{fig:spectra_comparison}
\end{figure}

\subsection{Spectral Comparison}
\label{spectra_comparison}

Figure \ref{fig:spectra_comparison} compares the spectra of SN 2025qe with those of the intermediate-luminosity SNe Iax SN 2019muj and SN 2024pxl at similar phases. The comparison spectra of SN 2024pxl and SN 2019muj were obtained from the Weizmann Interactive Supernova Data Repository(WISeREP\footnote{\url{https://www.wiserep.org/}}; \citealt{2012PASP..124..668Y}). Overall, the spectral evolution of SN 2025qe closely follows that of both comparison objects, reinforcing its classification as an intermediate-luminosity member of the SN Iax subclass. Nevertheless, some notable differences are present during the early phases.

At the earliest epoch ($\sim$ -2~day), all three objects exhibit prominent Fe group absorption features together with weak to moderate IME signatures. The Fe\,{\sc iii} absorption near 4300 \AA and the Fe\,{\sc ii}/Fe\,{\sc iii} complex around 5000 \AA are noticeably stronger in SN~2025qe than in either SN~2019muj or SN~2024pxl.  A similar trend is also observed near maximum light, where the Fe group absorption complexes remain stronger than in the comparison objects. In contrast, the IME features of SN 2025qe are broadly comparable to those seen in other intermediate luminosity SNe Iax. The Si\,{\sc ii} $\lambda$6355 absorption is clearly detected at early phases and exhibits a strength similar to that observed in SN~2019muj. Likewise, the C\,{\sc ii} $\lambda$6580 feature is present before and around maximum light, resembling the behavior seen in both SN~2019muj and SN~2024pxl. 

The spectral similarities become even more apparent by approximately one week after maximum. At phases of $+$8 to $+$9 d, the spectra of all three objects are dominated by Fe\,{\sc ii} absorption features, while the Si\,{\sc ii} and C\,{\sc ii} lines weaken considerably. The overall morphology of SN 2025qe closely resembles that of SN 2019muj and SN 2024pxl, including the development of a prominent Ca\,{\sc ii} near-infrared triplet. By $+$26 d, the spectra have evolved into the characteristic Fe-group dominated appearance of SNe Iax, with broad Fe\,{\sc ii} absorption complexes shaping much of the optical spectrum and the Ca\,{\sc ii} near-infrared triplet remaining one of the strongest features in all three objects. 

The stronger Fe-group absorption observed in SN 2025qe near maximum light may indicate differences in the distribution of iron-group material, ejecta density structure, or ionization conditions relative to other intermediate-luminosity SNe Iax. In particular, the higher density of Fe-group transitions can increase the line opacity and enhance the redistribution of radiation toward longer wavelengths. This may be related to the wavelength-dependent broadening seen in the post-maximum light curves (See Section \ref{lc_comparison}), although additional effects such as differences in ejecta density, composition, and radioactive-energy deposition may also contribute. Despite these differences, the convergence of the post-maximum spectra demonstrates that SN 2025qe shares the same overall spectroscopic evolution as other intermediate-luminosity members of the SN Iax class.

\begin{figure*}
    \centering
     {\includegraphics[width=0.8\textwidth]{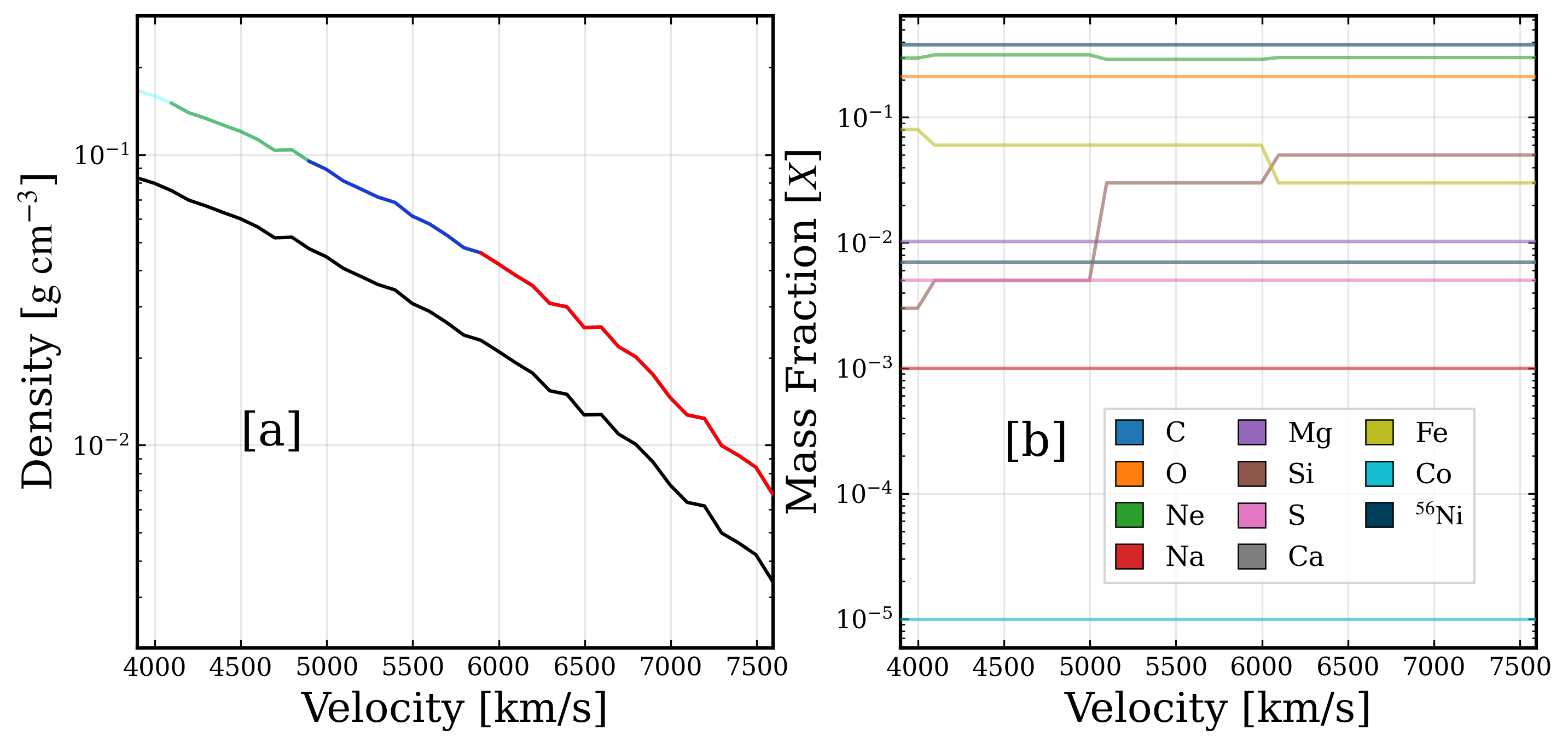}}
    \caption{[a] Density profile used for the \textsc{tardis} models. The different colors in panel [a] points to the density of the ejecta for each epoch. The density profile is the same but extend inwards  in velocity space - $v_{inner}=$ 5897 (red), 4897 (blue), 4098 (green), 3898 (cyan) $\rm km~s^{-1}$, 
    The black solid line indicates the angle-averaged density profile of the model `def$\_$r60$\_$d2.6$\_$Z'. [b] Mass fractions of the different elements used for the simulations. The density and mass fractions are defined at $\sim$ 100 secs from the explosion. We note that Ca and C have same mass fractions of X=0.007 and hence indistinguishable in panel [b].}
    \label{fig:tardis_dens_composition}
\end{figure*}

\begin{figure*}
    \centering
    {\includegraphics[width=0.8\textwidth]{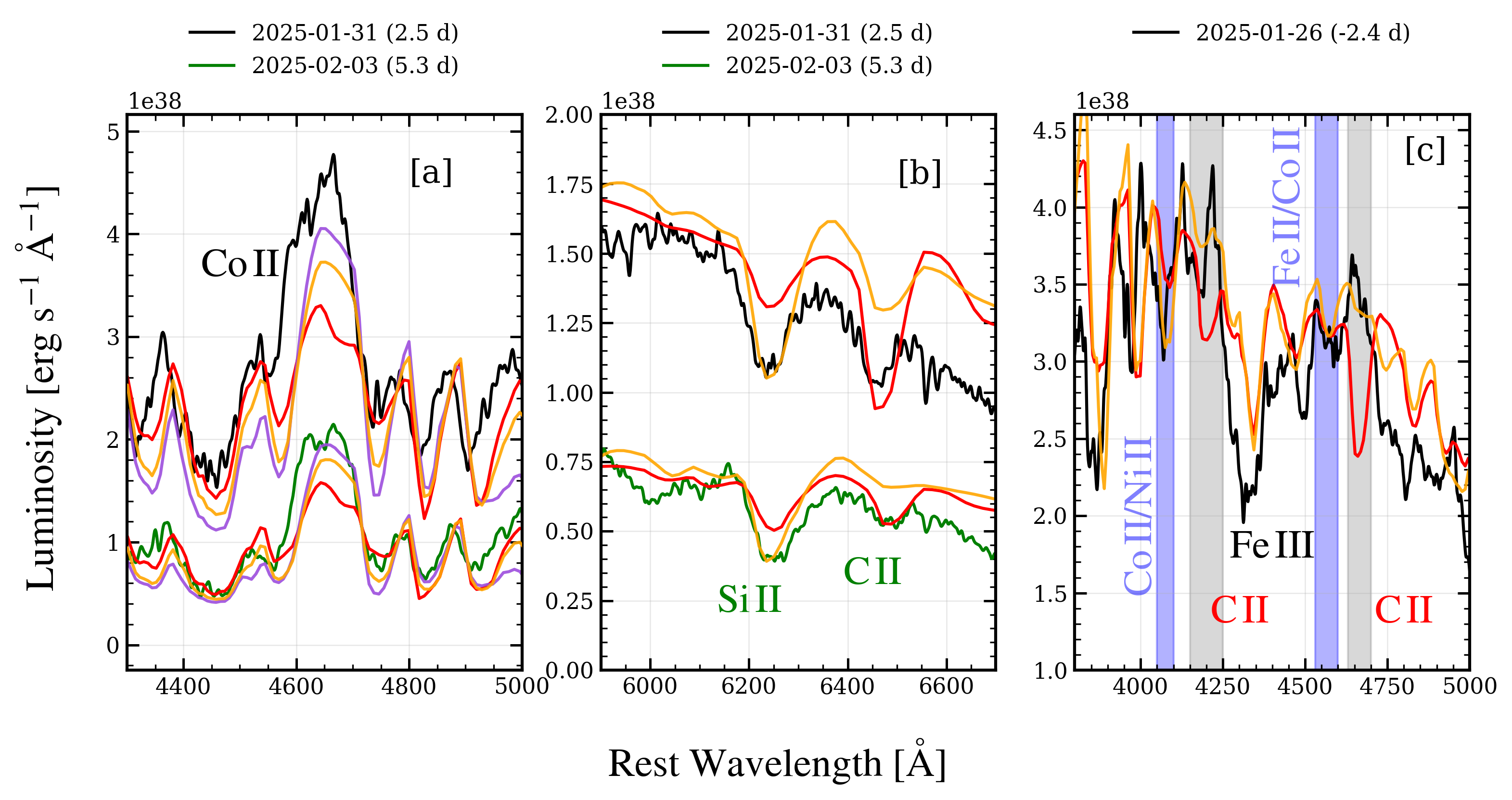}}
    \caption{The model spectra shown in yellow is with modified composition (Figure~\ref{fig:tardis_dens_composition}[b]) and density ($\times$2), while the red spectra are with \textsc{hemsa} density and composition. [a] Effect of varying the density on the line profile due to Co\,{\sc ii} shown for epochs 2.5~d and 5.3~d. The spectrum in magneta is base model density $\times$ 4. [b] Effect of using a stratified composition for Si at two different epochs. We see the effect of a stratified model more clearly in the later epoch. [c] Effect of changes in the density and composition for the pre-peak spectrum around the wavelength region between 4000 -- 5000 \AA.}
    \label{fig:tardis_model_features}
\end{figure*}

\subsection{Spectral modeling with TARDIS}  \label{spectral_modeling}

Most forward modeling techniques like \cite{2014MNRAS.438.1762F}, \cite{2022A&A...658A.179L} explore a range of parameters like the number of ignition spots (multi-spot or single-spot), distance of the spot from the center, metallicity, and rotation to explain the class of 2002cx-like SNe. However, individual SN radiative transfer studies focus on building self-consistent solutions to map the ejecta responsible for forming the spectral features \citep[see][]{2005MNRAS.360.1231S, 2014MNRAS.445..711S, 2024MNRAS.529.3838A}. 
To understand the conditions of the ejecta, such as the temperature and density profile, and to identify spectral features in the early phase, we use the radiative transfer code \textsc{tardis} \citep[][v2025.08.17]{2014MNRAS.440..387K, kerzendorf_2025_16888341}. 
The code takes a density and radially varying or uniform composition profile to calculate a spectrum for a given epoch. In this work, we have used the \texttt{nebular} mode for ionization, \texttt{dilute-lte} for excitation, radiative rates are obtained using \texttt{dilute-blackbody}, and the line interactions are treated in \texttt{macroatom} scheme.

\input{tardis_model}

We have used the angle averaged density profile of the model `def$\_$r60$\_$d2.6$\_$Z' (alternatively referred as the `base model') from \cite{2022A&A...658A.179L}. In this model a single spot ignition is initiated at a distance 60 km off-center, with a central mass density of 2.6 $\times$ 10$^{9}$ $\rm g~cm^{-3}$ and with a solar metallicity. Our choice of this model is motivated by the similarity of the model's $^{56}$Ni (0.018 \Msun) mass in the ejecta to the analytic model fitting to the observed light curves (0.02 \Msun). The model has an ejecta mass of 0.05 \Msun~ and a kinetic energy of 0.082 $\times$ 10$^{50}$ erg. 

We note that we use this to narrow down our searches for an appropriate density profile. However, we need to modify the density profile as discussed in the following. The angle averaged composition profile (mass fractions as a function of velocity) of this model is `mixed' in the sense that most elements considered for our spectral simulation are present at all velocities. Initially, for \textsc{tardis} models, we take the mean of the composition over the extent of the ejecta from the base model profile. We used $^{12}$C, $^{16}$O, $^{20}$Ne as the unburned species from the progenitor WD, $^{22}$Na, $^{24}$Mg, $^{28}$Si, $^{32}$S, $^{40}$Ca, $^{48}$Ti, $^{52}$Cr as intermediate mass elements (IMEs). We added stable Fe ($^{54}$Fe, $^{56}$Fe, $^{57}$Fe, and $^{58}$Fe), stable Co ($^{59}$Co), and $^{56}$Ni as the iron group elements (IGEs). The mass fraction of the radioactive isotope $^{56}$Ni is defined at 100 seconds from explosion (taken from \textsc{hesma}), and \textsc{tardis} allows for the decay of $^{56}$Ni. 

Initially, we model our observed spectra using similar density and composition for the model `def$\_$r60$\_$d2.6$\_$Z' and provide time since explosion ($t_{exp}$), velocity of an inner boundary ($v_{inner}$), an emergent luminosity of the supernova at the given epoch ($L_{SN}$), the temperature of the innermost boundary ($T_{inner}$). So, the features that are formed are above this $v_{inner}$ and may not be similar to the \textsc{hesma} model spectrum at the given epoch. There are degeneracies between some of these parameters and an equally good model may be produced by some other choice of $v_{inner}$, $t_{exp}$, $T_{inner}$, density profile and composition values \citep[see for example][]{2021ApJ...916L..14O, 2026ApJ..1002L..11L}. Here, we used our observations to reduce the number of free parameters. For example, the measured Si\,{\sc ii} $\lambda6355$ velocity is $\sim$6200 km~$\rm s^{-1}$ around $-$2.4 day. The time since explosion and also the luminosity (taking into account distance and reddening) are fixed.

In this set of modeling, we notice that there are features that are stronger or weaker than the observed features. We find the C\,{\sc ii} $\lambda6580$, $\lambda7234$ are much stronger in the spectrum and C\,{\sc ii} $\lambda6580$ remain strong till 5.4 days since maximum, the Si\,{\sc ii} feature is weaker than the observations.

\begin{figure*}
\centering
\includegraphics[width=0.47\textwidth]{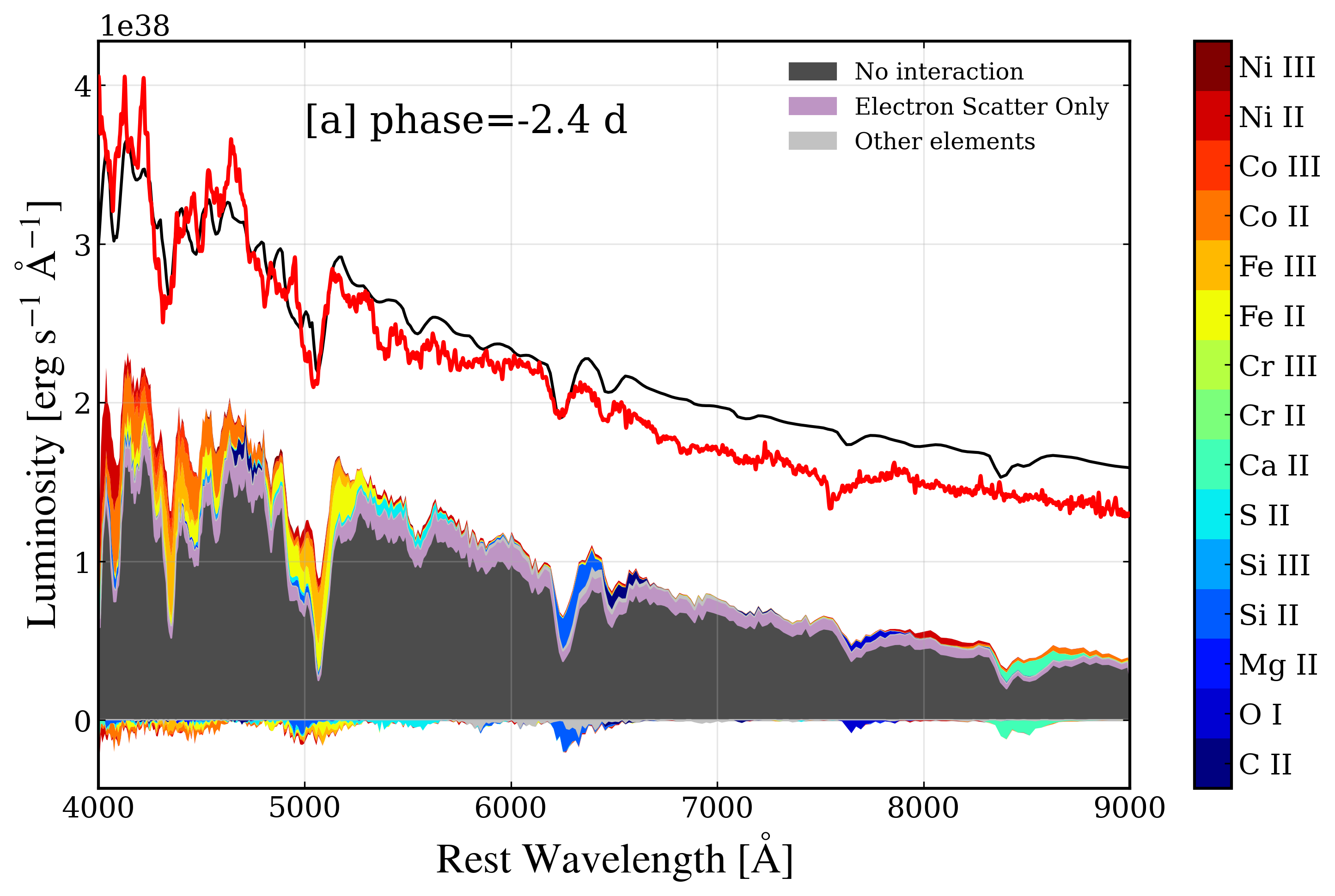}
\includegraphics[width=0.47\textwidth]{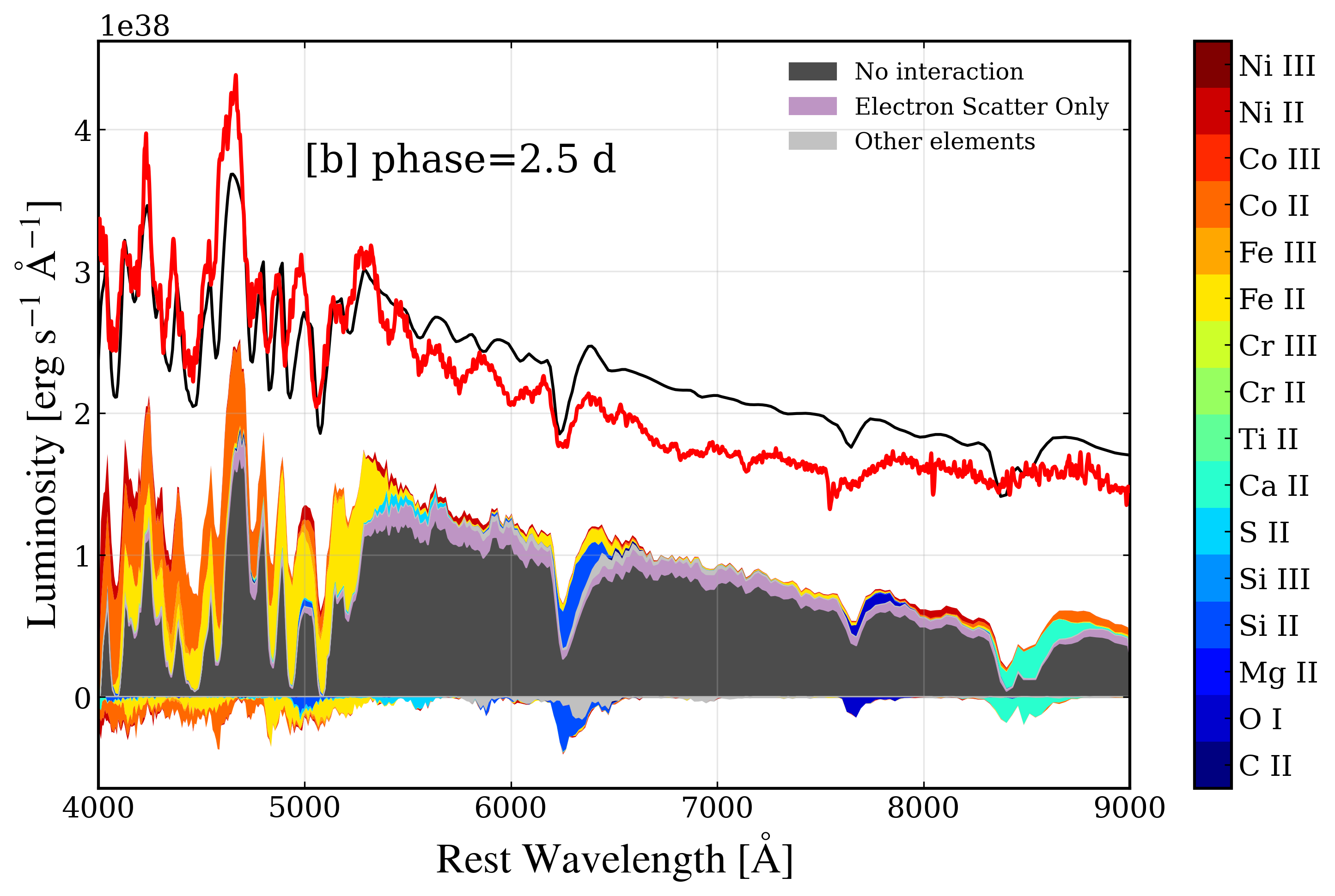}
\vspace{0.02cm}
\includegraphics[width=0.47\textwidth]{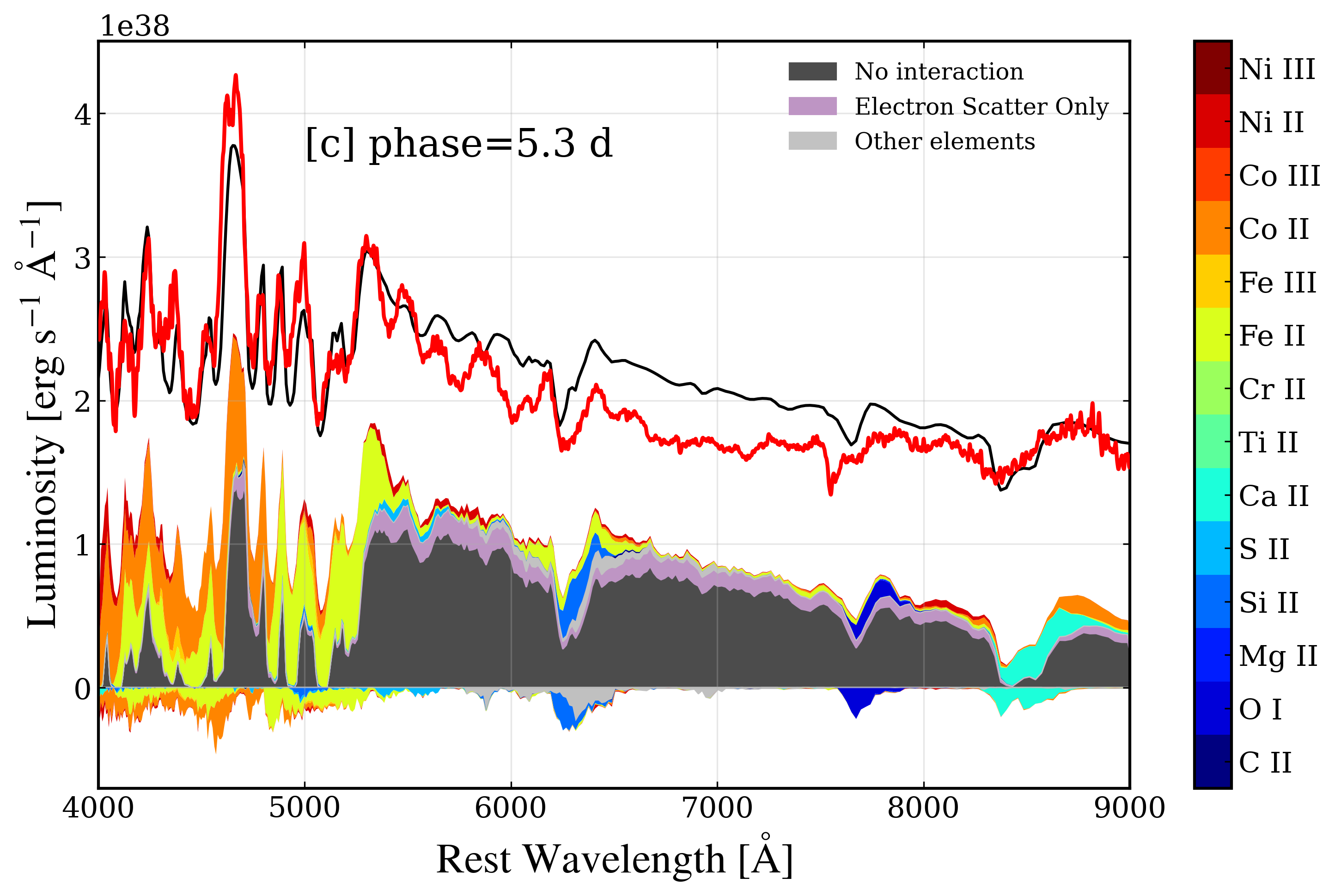}
\includegraphics[width=0.47\textwidth]{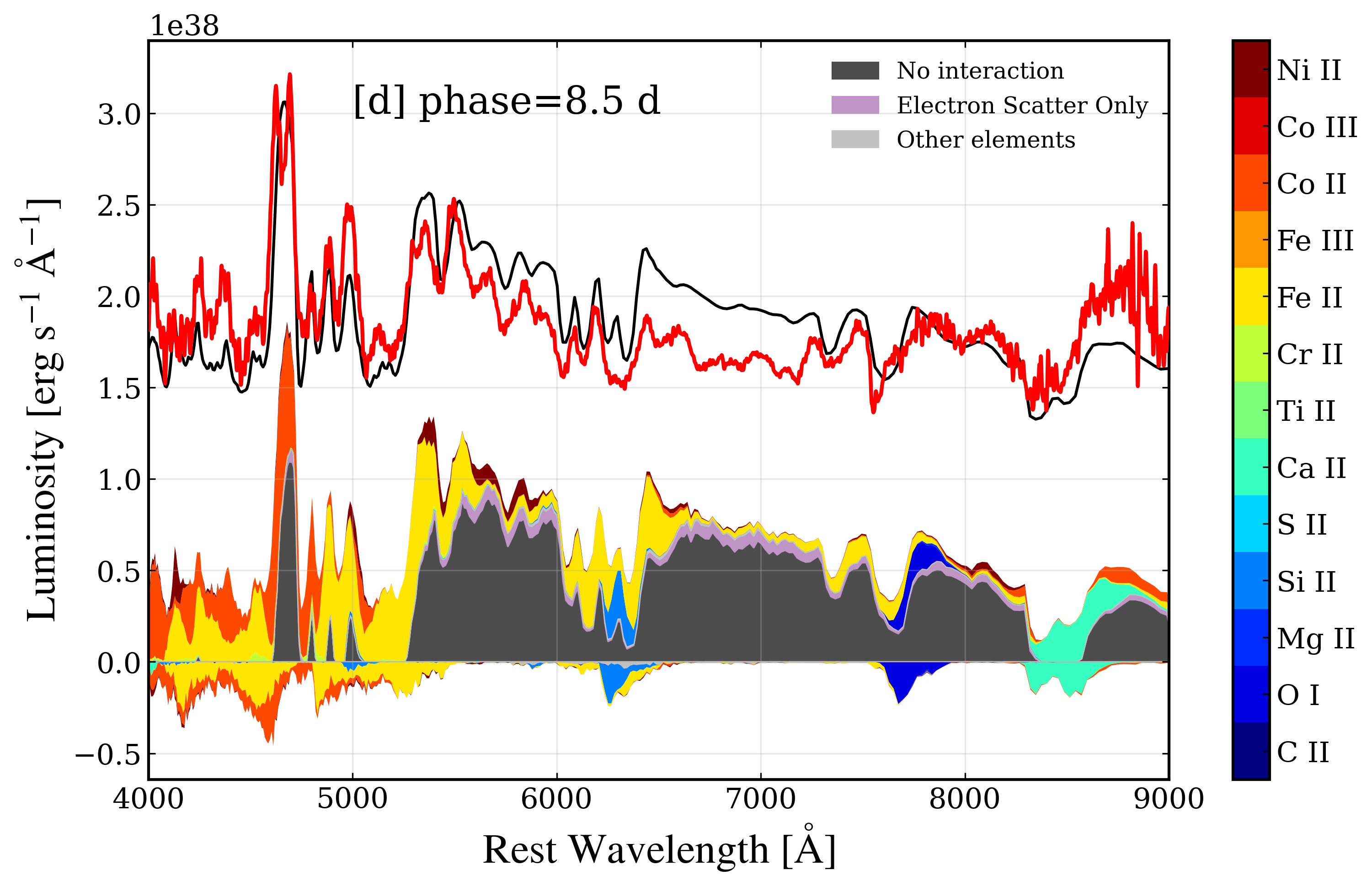}
\caption{Model spectra (black) generated with \textsc{tardis} overplotted with the observed ones (red). 
The rest frame phases are shown on each panel. We multiply both the observed and the model spectra by the same factor for visual clarity.
The Spectral Element Decomposition plot - showing the contributions of different ions to the spectrum. The individual panels illustrate how these ionic contributions evolve over time. Absorption is shown below zero on the $y$-axis, while emission is shown above zero. Electron scattering provides a prominent continuum till $\sim$5.3 days since peak.}
\label{fig: tardis_models_and_observations}
\end{figure*}

For modeling of the observed spectral features we have increased the density (base model $\times$ 2, see Figure~\ref{fig:tardis_dens_composition}[a]), which essentially means that we have a larger ejected mass than that in the model `def$\_$r60$\_$d2.6$\_$Z'. The modified model has a total mass of $\sim$0.04 $M_{\odot}$ above $\sim$ 3900 km~$\rm s^{-1}$ (see Table~\ref{tab:tardis_model_parameters}). However, this is much less than that found from the radiation diffusion fit to the pseudo-bolometric light curve. This is expected since the light curve fit uses a constant opacity and a characteristic expansion velocity of the ejecta. This means that it represents the ejecta mass with a single velocity, while much of the ejected mass could be below a certain inner velocity, and a fraction of the total mass is sufficient to form the spectral features.

This modification seems to better represent the feature around 4600 \AA~ as seen in the observed spectra. We show the comparison of the spectra with the models in Figure~\ref{fig:tardis_model_features} and Figure~\ref{fig: tardis_models_and_observations}. The increase in density provides more free electrons for recombination, and hence we find the model features to be consistent with the observations, particularly the feature due to Co \,{\sc ii} around 4600 \AA~ is getting stronger. We note that the stable Co present in our model is very small, but as $^{56}$Ni decays, the amount of $^{56}$Co increases and the feature attributed to a Co \,{\sc ii} becomes stronger (see Figure~\ref{fig:tardis_model_features}[a]). 

The density variation, combined with a uniform and constant Fe mass fraction, strengthens the Fe features more than in the observed spectrum. We therefore adopted a more stratified Fe to reproduce the blue region ($4000$--$5200,\text{\AA}$) of the early-phase spectrum. We also required a stratified Si that increases outwards to reproduce the Si\,{\sc ii} feature (see Figure~\ref{fig:tardis_dens_composition}[b] and Figure~\ref{fig:tardis_model_features}[b]). Note that our model uses a single value of $T_{inner}$ and no $^{56}$Ni heating the ejecta. Hence, we could not reproduce the continuum, especially in the red. However, the effect of a stratified and uniform Si and Fe, and the impact of increased density on the 4600 \AA~feature is discernable  (Figure~\ref{fig:tardis_model_features}). The feature near 4600 \AA~in our models become stronger with density (we increased it by 2$\times$ and 4$\times$), the asymmetric Si\,{\sc ii} feature around 6200 \AA~ can be represented well by a layered composition. 
Features due to C\,{\sc ii} are also observed around $\sim$4600 \AA~ and $\sim$4200 \AA~ which is not present in the observed spectrum, pointing towards an overall higher C mass fraction in the ejecta above $v_{inner}$ in the model without any modification (Figure~\ref{fig:tardis_model_features}[c]). 

We also generated the Spectral Element Decomposition (SDEC) plot, which serves as a basis for the line identifications during the early phase (Figure~\ref{fig: tardis_models_and_observations}). The major ions that contribute to the model spectra are C\,{\sc ii}, O\,{\sc i}, Si\,{\sc ii}, S\,{\sc ii}, Ca\,{\sc ii}, Cr\,{\sc ii}, Fe\,{\sc ii}, Fe\,{\sc iii}, Co\,{\sc ii}, Co\,{\sc iii}, Ni\,{\sc ii}. 
Early in the evolution ($\sim$2.4 days before $g$-band maximum), the spectrum shows a blend of Fe\,{\sc ii} and Fe\,{\sc iii} around 5000\,\AA, which becomes progressively dominated by Fe\,{\sc ii} after the peak.

The feature near 5400\,\AA~is initially dominated by S\,{\sc ii}, while Ni\,{\sc ii} becomes dominant by $\sim$8.6 days after maximum. The Co\,{\sc ii} feature near 4600\,\AA~strengthens relative to the continuum over time. The feature near 6200\,\AA, attributed to Si\,{\sc ii} $\lambda6355$, becomes blended with Fe\,{\sc ii} approximately 2.5 days after peak. At redder wavelengths, Ca\,{\sc ii} is blended with Co\,{\sc ii}.

The transition from Fe\,{\sc iii} to Fe\,{\sc ii}, and Co\,{\sc iii} to Co\,{\sc ii} can be understood from the fact that there is sufficient ejected mass but the temperature drops as a result of low $^{56}$Ni production. In a typical thermonuclear explosion, an increase in $^{56}$Ni increases the radiation temperature, and thus would produce more ionization. However, a lower $^{56}$Ni, together with the expansion of the ejecta, would make the explosion cool faster and therefore there would be a change in the radiation field and the ionization state. A low amount of $^{56}$Ni would come from weak explosions that produce less nuclear burning, less ejected mass and are less energetic. However, ignition geometry and the amount the WD expands due to the initial explosion can make the relation between ejected mass and $^{56}$Ni non-linear. We consider a layered composition for stable Fe, Si while all the other elements are mixed throughout. The mass fraction needs to be normalized to 1, so the effect of normalization can be seen in Ne. Although stable Fe is produced as a result of CO burning, the layered composition can indicate an aspherical explosion.

\subsection{Late Phase Spectral modeling with \textsc{syn++}}
\label{syn++}

\begin{figure}
	
    \centering
	{\includegraphics[width=\linewidth]{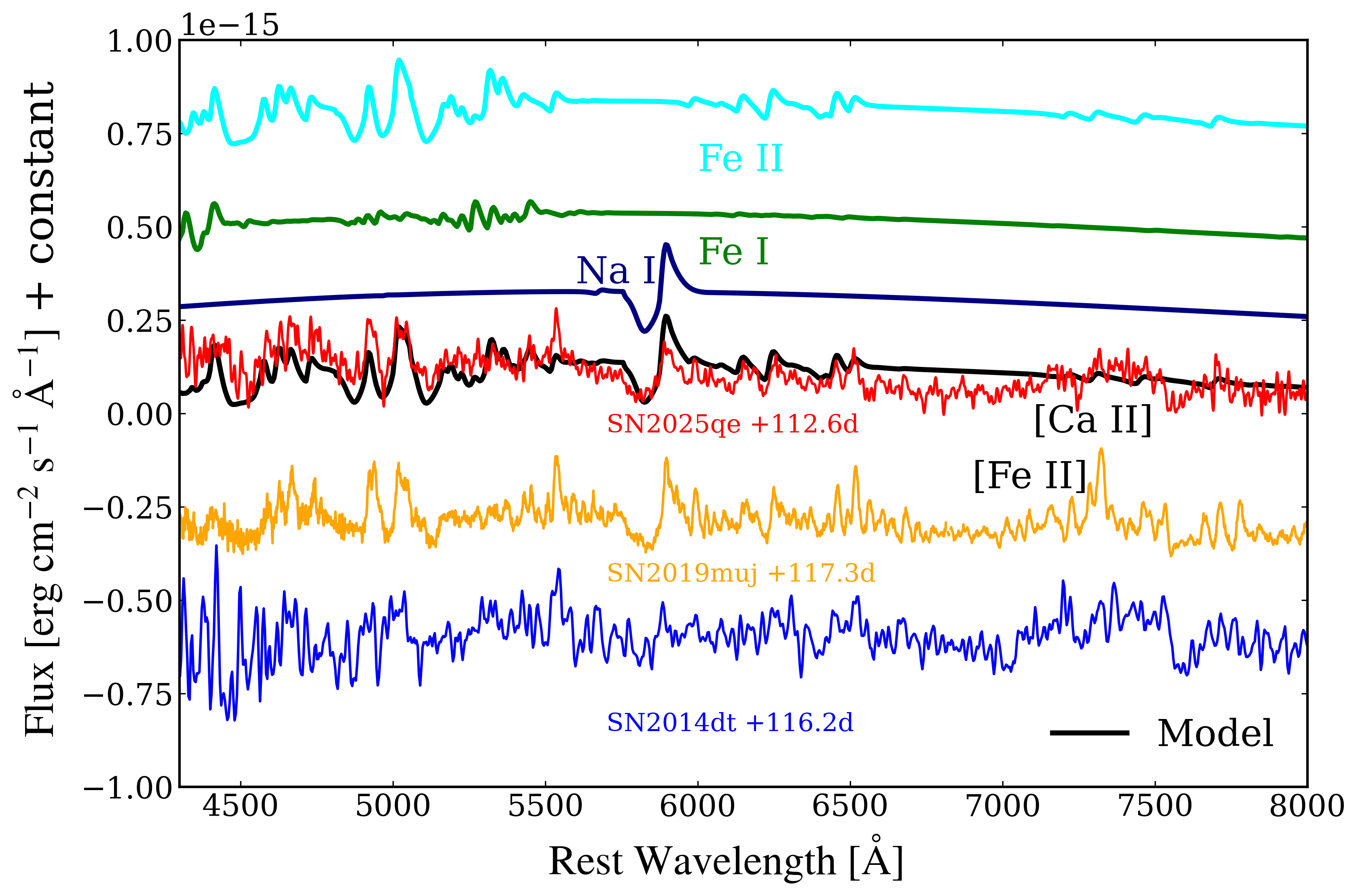}}
	
	\caption{\texttt{syn++} models plotted along with the observed spectrum at $\sim$112 days. We separately show the ions responsible for the formation of the spectrum under the assumptions that a sharply defined photosphere is still valid. The observed spectrum is redshifted corrected and dereddened. We also show the spectrum of SN2019muj and SN2014dt around similar phases for comparison. The comparison spectra have been redshift corrected. We also show the identification of [Ca\,{\sc ii}] and [Fe\,{\sc ii}] features in SN2019muj adapted from \cite{2021PASJ...73.1295K}.}
	\label{fig:syn++}
\end{figure}

We perform line-identification study of the spectrum observed at $+$112.6~d since maximum with \texttt{syn++} (Figure \ref{fig:syn++}). \texttt{syn++} is a parametrized code \citep{2011PASP..123..237T} that can be used to identify lines in an expanding homologous ejecta of a supernova. Features are identified in an empirical spirit by taking into account multiple resonant scattering. The ejecta is illuminated with blackbody radiation above an opaque photosphere at temperature $T_{ph}$, and velocity $v_{ph}$. An exponential density profile is chosen with an e-folding length of the profile given by $aux$. Lines are treated under Sobolev approximation and a reference optical depth $log\_{\tau}$ is set at $v_{ref}$, and the optical depth is calculated assuming an exponential form. Typically lines are formed at the photosphere, but the line may be detached, so $v_{min}$ and $v_{max}$ controls the extent of the ejecta where lines are formed in velocity space. \texttt{syn++} assumes that the ions follow the Boltzmann excitation given by $T_{exe}$. 

We used \texttt{syn++} qualitatively to find whether the spectrum at $+$112.6 d can still show superimposing P-Cygni features. \texttt{syn++} has been used till $\sim$100 days for exploring the late phase spectral line identifications in SN 2014J \citep{2016MNRAS.460.1614V}. Since the assumptions of \texttt{syn++} may not hold at these phases, we do not derive any physical parameters from the synthetic model, and we caution the reader that a detailed radiative transfer model treating non-thermal energy deposition, ionization and excitation, including forbidden transitions, would be required. However, from our fit we see that the synthetic spectrum reproduces the Na\,{\sc i} feature at 5900 \AA~and Fe\,{\sc ii} features are present between 6200 \AA~and 6600 \AA. In the lower wavelength region also we find features in the observed spectrum that can be identified with Fe\,{\sc ii} around 5000 \AA. To get a better fit we also used Fe\,{\sc i}, however the line strengths are weak at this phase in this wavelength region. Overall, a decent fit to the spectrum at this phase may be possible because a photospheric assumption may still hold for the ejecta of SNe Iax, as they are known to exhibit photospheric and permitted-line dominated spectra to unusually late epochs \citep{2023ApJ...951...67C}. The model $v_{ph}$ is set at 1000~$\rm km~s^{-1}$, $v_{ref}$ at 5000 $\rm km~s^{-1}$, $v_{max}$ is placed at 7000~$\rm km~s^{-1}$.
The photospheric temperature is at 4000K and the Boltzmann excitation temperature at 5000 K. The \texttt{syn++} fit suggests that the ejecta is still dense enough to form the permitted P-Cygni features.

Figure \ref{fig:syn++} also shows the comparison of the $+$112.6~d spectrum with two other SNe Iax, SN~2019muj and SN~2014dt around similar phases. The relatively weak development of the forbidden [Ca\,{\sc ii}] feature in SN~2025qe compared with SN~2019muj and SN~2014dt indicates that the line-forming region in SN~2025qe may be comparatively denser. High densities can suppress the development of forbidden transitions while permitted Fe-group features remain prominent \citep{2022ApJ...941...15M}. 

\section{Discussion}
\label{discussion}

The photometric and spectral evolution of SN~2025qe shows similarities with other SNe Iax. The observed peak $g$-band absolute magnitude of $M_{g} \approx -16.16\pm0.15$~mag places it among the intermediate luminosity Iax systems such as SN 2019muj and SN 2024pxl. 
Analytical model fits to the pseudobolometric light curve of SN 2025qe indicate a $^{56}$Ni mass of $0.02\,M_\odot$, within the range observed for other SNe Iax. Modeling the spectra using the radiative transfer code \textsc{tardis} suggests that a density enhancement and stratified composition relative to the three-dimensional deflagration model `def$\_$r60$\_$d2.6$\_$Z', which produces a \(^{56}\mathrm{Ni}\) mass consistent with that inferred for SN 2025qe, provides a good match to the observed spectra.

The \cite{2022A&A...658A.179L} deflagration models do not fully reproduce the observed multi-band light curves of SN~2025qe, particularly the broader and more slowly declining light curves at longer wavelengths. This discrepancy suggests that additional physical effects may be required to reproduce the observed photometric evolution. Radiative transfer calculations that include a central, luminous bound remnant, such as the models presented by \cite{2024MNRAS.530.1457C}, have been shown to provide a better match to observations, particularly in the redder bands.
 In this scenario, the high energy photons are trapped in the remnant which heats the remnant and emits like a blackbody. As time evolves the spectral energy distribution of the remnant shifts towards redder wavelengths, which helps prevent the light curves from declining faster as compared to models without any energy injection from the remnant.  

Another proposed model is an inner dense component with Fe-group rich composition that is distinct from the outer ejecta \citep{2022ApJ...941...15M}. In this proposed scenario, the energy input from the $^{56}$Ni decay chain within the remnant of the WD can unbind a significant fraction of the remnant when the decay energy increases in comparison to the binding energy. In such a case, the total luminosity is powered by the energy deposited by the WD bound remnant, unbound denser ejecta and the initial ejecta from the explosion.

\begin{figure*}
    \centering
    \resizebox{\hsize}{!}{\includegraphics{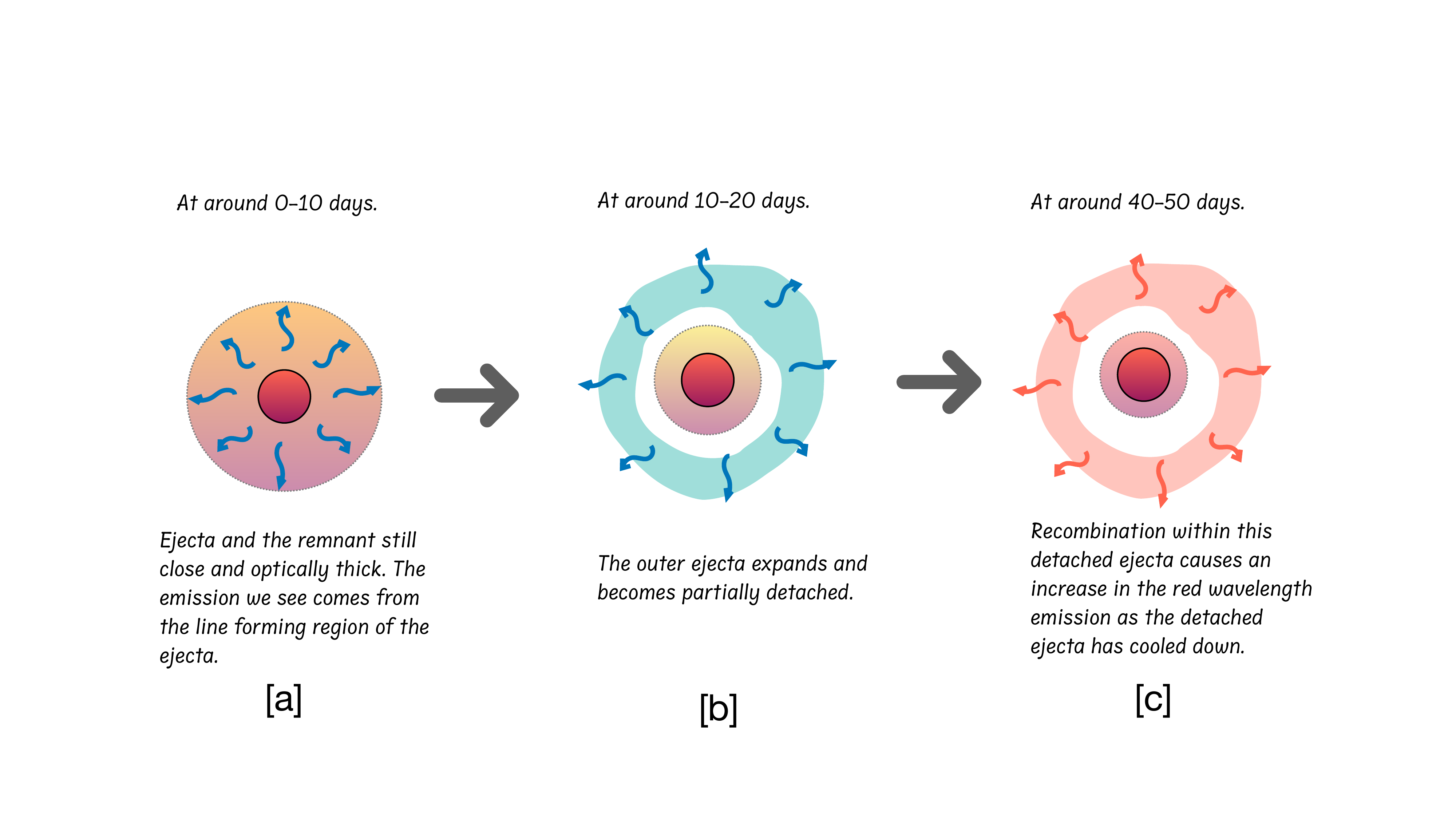}}
    \caption{Schematic time evolution of SN 2025qe. The visualization is generated with Apple-Keynote.}
    \label{fig:2025qe_possible_explanation}
\end{figure*}

Based on these two models, we suggest a possible qualitative scenario for the ejecta structure of SN 2025qe that could account for some of its distinctive observational properties. We consider a two-component ejecta configuration consisting of a distinct outer shell, separated in velocity space from an inner region possibly containing a bound remnant and denser inner ejecta. A schematic representation of this possible configuration is shown in Figure~\ref{fig:2025qe_possible_explanation}.

This scenario can provide a possible explanation of the slower decline of flux in redder bands, as seen around $\sim$ 40 days in SN 2025qe.  Around $\sim$ 10 days from the explosion, the higher optical depth causes the emergent radiation to originate from an extended line-forming region. Thus, it prevents the components from being distinguished. This extended line-forming region may contain a relatively dense and Fe-group-rich component, which could account for the stronger Fe-group absorption features observed in the near-maximum spectra of SN~2025qe (See Section \ref{spectra_comparison}) and can contribute to the redistribution of flux towards longer wavelengths through increased line opacity and fluorescence. 

A smooth, homologously expanding ejecta with a rapidly declining outer density profile would result in rapidly decreasing optical depth. Hence, we consider a density-enhanced outer shell to sustain sufficient optical depth at later times. As the radius increases and the optical depth decreases, the contribution of the density-enhanced outer shell becomes increasingly important. This density enhancement in the outer shell may possess an asymmetrical structure and can help sustain the redistributed flux at redder wavelengths at later phases. 

The high energy $\gamma$-rays, $X$-rays and leptons ($e^{-}$/$e^{+}$) from radioactive decays, emitted from the inner region, continue to deposit fraction of their energy in the outer shell, where it is re-emitted through UV, optical and infrared radiation. After formation of the shell, it remains dense enough to redistribute photons of shorter wavelengths into longer wavelengths. 
If the mass of $^{56}$Ni present in this shell is not significant enough to cause local heating of the ejecta, the adiabatic expansion of the shell will cause cooling more rapidly compared to the inner region. The lower temperature favors recombination to lower ionization states, while fluorescence redistributes the UV and optical photons to longer wavelengths, thus producing a sustained red band emission till late phases.  

The weaker forbidden [Ca\,{\sc ii}] emission in SN~2025qe, as noted in Section \ref{syn++} can also be qualitatively explained with a dense component persisting to late epochs. In the proposed picture, such a dense region could be associated with the inner ejecta and/or with the density-enhanced outer component, depending on the detailed distribution of the material in velocity space.

The scenario outlined above is purely qualitative and is not supported by quantitative radiative-transfer or hydrodynamic modeling that is beyond the scope of this work. 

Though SN~2025qe lies in the intermediate-luminosity regime of SNe~Iax, its photometric and spectroscopic evolution shows considerable diversity even among events of similar luminosity. Its wavelength-dependent light-curve broadening, Fe-group absorption, and late-time spectral properties suggest that variations in ejecta composition, density, and structure may play an important role in shaping the observed diversity. A larger sample of well-observed intermediate-luminosity SNe~Iax, together with detailed multidimensional modeling, is therefore needed to understand the diversity of ejecta structures and other observables.

\section{acknowledgments}

\noindent This research was supported by the Science and Engineering Research Board (SERB), Department of Science and Technology, Government of India, under the project ``Understanding the progenitor and explosion mechanism of supernovae through studies in the nearby Universe" (grant number CRG/2022/007688).

H.D. received support through the CRG/2022/007688 grant. G.C.A. is supported by the Senior Scientist Programme of the Indian National Science Academy (INSA). A.D. is supported by the National Science and Technology Council of Taiwan (NSTC grant 115-2123-M-008-001).

We are grateful to the staff at the Indian Astronomical Observatory (IAO), Hanle, and the CREST campus, Hosakote, for their invaluable support in facilitating observations with the Himalayan Chandra Telescope (HCT). The GROWTH-India Telescope (GIT; Kumar et al 2022) is a 70 cm telescope with a 0.7 deg field of view set up by the Indian Institute of Astrophysics (IIA) and the Indian Institute of Technology Bombay (IITB) with funding from the Indo-US Science and Technology Forum and the Science and Engineering Research Board, Department of Science and Technology, Government of India. It is located at the Indian Astronomical Observatory (IAO, Hanle). We acknowledge funding by the IITB alumni batch of 1994, which partially supports the operation of the telescope. Telescope technical details are available at https://sites.google.com/view/growthindia/. We also extend our appreciation to the HCT and GIT observers who conducted Target of Opportunity (ToO) observations as part of the early follow-up efforts.

This study utilized resources from the NASA Astrophysics Data System (ADS)\footnote{\url{https://ui.adsabs.harvard.edu/}} and the NASA/IPAC Extragalactic Database (NED). We also thank the Weizmann Interactive Supernova Data REPository (WISeREP) for their data services. Photometric measurements were obtained from the \textit{Swift}/UVOT archive and the Zwicky Transient Facility (ZTF) archive via Lasair\footnote{\url{https://lasair-lsst.lsst.ac.uk/}} \citep{2019RNAAS...3...26S}.

We acknowledge the use of \textsc{tardis}, a community-driven software package for supernova spectral analysis \citep{2014MNRAS.440..387K, kerzendorf_2025_16888341}. Development of \textsc{tardis} was supported by GitHub, the Google Summer of Code, and ESA's Summer of Code in Space programme. This project is fiscally sponsored by NumFOCUS and relies heavily on Astropy. Our simulations used the \texttt{TARDIS} atomic data file \texttt{kurucz\_cd23\_chianti\_H\_He\_latest.h5}.

We also accessed the Heidelberg Supernova Model Archive (\textsc{hesma}\footnote{\url{https://hesma.h-its.org/}}; \citealt{2017MmSAI..88..312K}). The following software and packages were employed in this work: \texttt{IRAF} \citep{tody1993iraf}, \texttt{PyRAF} \citep{2012ascl.soft07011S}, \texttt{NumPy} \citep{2011CSE....13b..22V}, \texttt{Matplotlib} \citep{2007CSE.....9...90H}, \texttt{SciPy} \citep{2020NatMe..17..261V}, \texttt{pandas} \citep{pandas}, \texttt{Astropy} \citep{2013A&A...558A..33A, 2018AJ....156..123A, 2022ApJ...935..167A}, \texttt{syn++} \citep{2011PASP..123..237T}, \texttt{emcee} \citep{2013ascl.soft03002F}, \texttt{corner} \citep{2016JOSS....1...24F}, \texttt{sncosmo} \citep{2016ascl.soft11017B}, and \texttt{extinction} \citep{2021ascl.soft02026B}.

\bibliography{ms}{}
\bibliographystyle{aasjournal}

\end{document}

%% file: Bolometry_Comparison_Table.tex
\begin{table}[htbp]
  \centering
  \footnotesize
  \setlength{\tabcolsep}{3pt}
  \caption{Comparison of peak $Ugriz$ pseudo-bolometric luminosity, $^{56}$Ni mass and other explosion parameters like the photon diffusion timescale ($t_{\rm lc}$), Ejecta Mass ($M_{\text{ej}}$) and Kinetic Energy of the Ejecta ($E_{\text{k}}$) of SN 2025qe with other Type Iax.}
  \label{tab:bolometry_comparison}
  \begin{tabular}{cccccc}
    \hline
    \hline
    Object & $L^{Ugriz}_{\text{max}}$ & $M_{\mathrm{Ni}}$ & $t_{\text{lc}}$ & $M_{\text{ej}}$ & $E_{\text{k}}$ \\
     & ($10^{41}~\text{erg}~{\text{s}}^{-1}$) & ($M_{\odot}$) & (days) & ($M_{\odot}$) & ($10^{50}~\text{erg}$) \\
    \hline
    SN 2025qe  & 5.69 & 0.02 & 10.69 & 0.45 & 0.70\\
    SN 2019muj & 6.04 & 0.02 & 6.22  & 0.13 & 0.16\\
    SN 2024pxl & 8.28 & 0.03 & 11.96 & 0.42 & 0.36\\
    SN 2020kyg & 1.71 & 0.005 & 6.36 & 0.15 & 0.18\\
    SN 2022eyw & 25.2 & 0.11 & 12.59 & 0.79 & 1.90\\
    \hline
  \end{tabular}
\end{table}

%% file: tardis_model.tex
\begin{table*}[!t]
\centering
\begin{threeparttable}
\caption{\textsc{tardis} model parameters}
\label{tab:tardis_model_parameters}
\begin{tabular}{ccccccc}
\hline
Model & Luminosity (erg $\rm s^{-1}$) & $t_{\rm exp}$ (days) & Phase (days)* & $T_{\rm inner}$ (K) & $v_{\rm inner}$ (km\,s$^{-1}$) & mass ($M_{\odot})$**\\
\hline
1 & 7.5 $\times$ 10$^{41}$ & 9.0 & $-$2.4 & 9873 & 5896 & 0.011\\
2 & 8.22 $\times$ 10$^{41}$  & 14.5 & 2.5 & 8352 & 4897 &  0.023\\
3 & 7.5 $\times$ 10$^{41}$ & 17.5 & 5.4 & 8162 & 4097 &  0.035  \\
4 & 5.43 $\times$ 10$^{41}$ & 21.0 & 8.6 & 6968 & 3897 & 0.038 \\
\hline
\end{tabular}
\begin{tablenotes}
\footnotesize
\item The outer boundary velocity, $v_{\rm outer}$ = 7596 km\,s$^{-1}$, is fixed for all models.
\item *Phase since $g$-band maximum.
\item **Mass contained within $v_{inner}$ and $v_{outer}$
\end{tablenotes}
\end{threeparttable}
\end{table*}